\documentclass[11pt]{article}

\usepackage[letterpaper,top=2cm,bottom=2cm,left=3cm,right=3cm,marginparwidth=1.75cm]{geometry}
\usepackage[T1]{fontenc}
\usepackage{lmodern}
\usepackage{microtype}
\usepackage{csquotes}

\usepackage{amsmath}
\usepackage{amssymb}
\usepackage{amsthm}
\usepackage{mathtools} % Extends amsmath
\usepackage{thmtools}
\usepackage{thm-restate} % Useful for restating theorems in the appendix
\usepackage{dsfont}

\usepackage{graphicx}
\usepackage{booktabs}
\usepackage[ruled, vlined]{algorithm2e}

\SetAlFnt{\small\linespread{1.1}\selectfont} 
\SetAlCapFnt{\small}
\SetAlCapNameFnt{\small}
\SetAlCapHSkip{0pt}
\IncMargin{-\parindent}
\SetAlgoInsideSkip{smallskip}

\usepackage[dvipsnames]{xcolor}
\usepackage[
    colorlinks=true,
    allcolors=blue,
    bookmarks=true,
    bookmarksopen=true,
    bookmarksdepth=3
]{hyperref}
\usepackage[nameinlink,capitalize]{cleveref} % Must be loaded after hyperref
\crefname{algocf}{algorithm}{algorithms}
\Crefname{algocf}{Algorithm}{Algorithms}

\newtheorem{theorem}{Theorem}
\newtheorem{corollary}[theorem]{Corollary}

\newtheorem{lemma}{Lemma}[section]
\newtheorem{proposition}{Proposition}[section]
\newtheorem{definition}{Definition}[section]

\newtheorem{remark}{Remark}[section]

\newtheorem*{itheorem}{Informal Theorem}

\newtheorem*{ktheorem}{Known Theorem}

\title{Online Block Packing and Multidimensional EIP-1559}

\author{Ariel Ben Eliezer\thanks{Hebrew University of Jerusalem.} \and Noam Nisan\footnotemark[1]}

\date{June 2026}

\begin{document}

\maketitle

\begin{abstract}
We consider the online algorithmic challenge that is faced by blockchains that have multidimensional block constraints and serve quasi-patient bidders. We first provide online approximation algorithms for the important special cases of small transactions or a small number of dimensions; this solves open problems left by [Babaioff and Nisan, EC 2025]. 
Second, we study multidimensional variants of Ethereum's EIP-1559 protocol. We show that if the block builders manage to approximately optimize each block's welfare myopically, then an approximation to the global offline optimal welfare is obtained. On the other hand, we show that, unlike in the single-dimensional case, EIP-1559 by itself does not guarantee any good approximation.
\end{abstract}

\section{Introduction}

\subsection{Background: Blockchains}

Blockchain platforms such as Bitcoin~\cite{bitcoin} and Ethereum~\cite{eth} maintain a public ledger in a decentralized manner, requiring trust not in any single entity, but in the honesty of a majority (or super-majority) of participants.
These systems operate by batching user-submitted transactions into sequentially produced blocks, for example, every 10 minutes in expectation in Bitcoin and every 12 seconds in Ethereum.
Each block's capacity is limited so that nodes with modest hardware can still participate.
Consequently, transactions compete for inclusion in blocks by offering fees.
Each blockchain implements a transaction fee protocol that determines which transactions are included, how much users pay, and who receives the payments.
Each transaction yields value to its creator upon inclusion in the blockchain. A major goal of
blockchain protocols is to maximize \enquote{social welfare}, the total value of executed transactions.
Because transactions arrive and blocks are produced over time, the resulting problem is inherently online.

To define the problem more precisely, we need to specify two key concepts: block capacity and transaction time sensitivity.
Simpler mechanisms, such as Bitcoin's, assign each transaction a single size and cap the total size per block.
More sophisticated systems may impose multidimensional constraints: each transaction consumes several resources, and each block has a limit on the total usage of every resource.
Ethereum, for example, defines two independent resources: \enquote{gas}~\cite{1559} and \enquote{blobs}~\cite{4844}, with some proposals~\cite{7706} for additional resource types.
Managing multiple resources has attracted growing attention~\cite{ADM25, DECA23, KKPZ25, LNPR25} because it models system constraints more accurately and can improve the utilization of the true system resources, at the cost of added complexity. 

In many practical scenarios, the value of a transaction to its user depends on the time of placing the transaction on the blockchain.
Transaction time sensitivity ranges from impatient bidders, whose transactions lose all value unless they are included in the next block, to fully patient bidders, whose transactions retain their full value regardless of delay.
A more general model involves \enquote{quasi-patient} users~\cite{PS25, GY24}, where the value decreases by some constant multiplicative discount rate for each unit of time delay.\footnote{An alternative model for intermediate time sensitivity would be to define a \enquote{deadline} for each transaction, where transactions do not lose any value until the deadline and then lose all of it. Our results also apply to this model.}

This paper provides two contributions. In the first part of the paper, we define and study the purely algorithmic online scheduling problem of packing transactions 
into multidimensional blocks so as to maximize social welfare. Motivated by the blockchain application, we term this problem \enquote{online block packing}. 
Our main result in
this part provides efficient online approximation algorithms for social welfare in the two most practically important cases: a small number of dimensions or small transaction sizes. 

In the second part, we focus on the multidimensional variants of Ethereum's EIP-1559 protocol and analyze 
their welfare guarantees. 
Our results here are nuanced: 
our main positive result is that if the block builder can myopically approximate the optimal welfare for each individual block among its allowed transactions, the combination of EIP-1559 and this builder successfully approximates the global offline optimum.
The main negative result, which we view as surprising, shows that the protocol's pricing mechanism alone does not guarantee a good approximation to social welfare.
Specifically, when paired with the most natural greedy block building algorithm, the approximation guarantee degrades by a factor proportional to the number of resources.

The remainder of this introduction is structured as follows. In \Cref{sub:online-block-packing}, we formally define the online block packing problem and present our pure algorithmic approximations and lower bounds. In \Cref{sec:1559}, we apply this theoretical framework to evaluate the multidimensional EIP-1559 protocol in practice, concluding with a summary table of all our established bounds.

\subsection{Online Block Packing} \label{sub:online-block-packing}

In the first part of this paper, we define and study the purely algorithmic online scheduling problem of packing quasi-patient transactions into multidimensional blocks to maximize social welfare. 
Babaioff and Nisan~\cite{BN25} studied this problem for the \emph{single-dimensional} case and \emph{fully patient bidders}. They showed that under these two restrictions,
a simple online algorithm can obtain near-optimal social welfare,
using mild resource augmentations termed \enquote{slackness and extension}. However, \cite{BN25} also established lower bounds showing that
\emph{every online algorithm} must lose at least a constant fraction of social welfare if either of these two restrictions is relaxed (even when similar resource augmentation is allowed).
They left open the question of whether any \emph{constant-factor} approximation is obtainable for either of these cases. We answer this question in the affirmative.

\subsubsection{The Model}

Consider the following setting:
Transactions arrive over time and must be packed, online, into successive blocks.
A transaction not selected for the current block remains in the \enquote{mempool} as a candidate for future blocks.
Thus, the process yields a series of multidimensional knapsack problems~\cite{KPP04}:
in each round \(t\), the algorithm packs a feasible subset of the outstanding transactions into the block,
while unscheduled transactions carry over
for future rounds, with their values appropriately discounted.
This problem shares conceptual similarities with several well-studied classes of online problems
such as online bin packing~\cite{CKPT17}, online scheduling~\cite{S98}, online packing and covering~\cite{BN09},
online vector bin packing~\cite{ACKS13}, online multiunit combinatorial auctions~\cite{BGN03}, and online multidimensional knapsack~\cite{YZHST21}, but is distinct from all of these and worthy of independent study, especially due to its blockchain applications.

We now define the formal setting. Transactions arrive dynamically over time and are indexed by a set $\mathcal{C}$. 
Each transaction $i \in \mathcal{C}$ is represented by a tuple $(a_i, v_i, \rho_i, w_i)$, where $a_i \in \mathbb{N}$ is the arrival time, $v_i \ge 0$ is the base value, $\rho_i \in [0,1]$ is the discount rate, and $w_i \in \mathbb{R}_+^m$ is the demand vector, with $w_{ij}$ denoting the amount of resource $j$ the transaction consumes. Executing transaction $i$ in block $t \ge a_i$ yields a discounted value of $v_i(1-\rho_i)^{t-a_i}$.

At each time step $t \in \{1, 2, \dots\}$, the algorithm constructs a \enquote{block} from the pool of available transactions (i.e., those with $a_i \le t$). Let the indicator variable $x^t_i \in \{0,1\}$ denote whether transaction $i$ is scheduled in block $t$. Every block is subject to a fixed capacity $B_j$ for each resource $j \in \{1, \dots, m\}$, imposing the constraint $\sum_{i \in \mathcal{C}} x^t_i w_{ij} \le B_j$. 
Since each transaction can be scheduled at most once, for every $i\in\mathcal{C}$ we must have $\sum_t x^t_i \le 1$.
The online nature of the problem dictates that the assignment to block $t$ depends solely on the information revealed up to that point.
The goal is to maximize the time-discounted total value, the \emph{social welfare}, up to some unknown horizon $T$.

\begin{definition}
	The social welfare of an allocation $x = \{x_i^t\}_{i\in\mathcal{C},t\in\mathbb{N}}$ from time $1$ to $T$ is
	defined to be $SW_{[1:T]}(x)=\sum_{t=1}^T \sum_{i\in\mathcal{C}} x^t_i v_i (1-\rho_i)^{t-a_i}$.
\end{definition}

This problem presents two fundamental challenges:
\begin{enumerate}
    \item \textbf{Computational Hardness:} Approximating the optimal solution for a single block, even within a factor of $1/m^{1/2-\epsilon}$, is NP-hard.\footnote{This is because Maximum Independent Set is equivalent to a special case of Set Packing, which itself is a special case of our single-block problem. The known hardness result for approximating Maximum Independent Set to within a factor of $n^{1-\epsilon}$ \cite{H99} translates directly to $m^{1/2-\epsilon}$ in our setting. For tight bounds, see~\cite{DKM24}.}
    \item \textbf{Online Nature:} The algorithm must schedule a block without having information about future transactions.
\end{enumerate}

Consequently, to obtain polynomial-time online algorithms, we must focus on settings where the single-block building problem is computationally tractable. There are two such basic cases that together cover most of the realistic scenarios for the blockchain application:
\begin{itemize}
\item \textbf{Small number of dimensions ($m$):} While in some cases we may have a large number of resources $m$, typically, $m$ is very small. For example, in Ethereum we have only two resources ($m=2$): \enquote{gas}~\cite{1559} and \enquote{blobs}~\cite{4844}.
\item \textbf{Small transaction sizes:} While some rare transactions may be large relative to block sizes, most are usually much smaller. Typically, the number of transactions that fit into a block is in the hundreds (Ethereum) or thousands (Bitcoin). Formally, we assume all transaction sizes are small relative to block capacities, meaning $q_{\max}:=\max_{i,j} \frac{w_{ij}}{B_j} \ll 1$. This assumption closely approximates the purely fractional packing problem (i.e., allowing transactions to be fractionally divided across multiple blocks with proportionally discounted values).
\end{itemize}

Finally, regarding \textbf{incentives}: this paper addresses the algorithmic problem and explicitly refrains from modeling users' strategic behavior. 
The reader should always keep in mind, however, that the algorithm runs within a strategic
context, which strongly influenced the design of EIP-1559. Formal modeling and analysis of these strategic aspects are beyond the scope of this paper and are left as important directions for future research.

\subsubsection{Algorithms for the Block Packing Problem}

Our main result shows that myopic optimization on a block-by-block basis results in a good approximation overall. Our main theorem considers any online algorithm that achieves at least a fraction $\lambda \le 1$ of the optimal possible welfare for the current block 
(myopically); we call such algorithms myopically-$\lambda$-optimal.

\begin{definition}
    An online algorithm is said to be \emph{myopically-$\lambda$-optimal} (for $\lambda \le 1$) if, at every time step $t$, the allocation it produces achieves at least a fraction $\lambda$ of the optimal possible welfare for that block, considering the set of currently available transactions (i.e., those that have arrived but remain unscheduled).
\end{definition}

Depending on the setting, this myopic benchmark can be defined with respect to either the optimal \emph{fractional} allocation or the optimal \emph{integral} allocation for the single block. We will explicitly state whether an algorithm is myopically-$\lambda$-optimal with respect to a fractional or integral benchmark when it is introduced. Regardless of the chosen benchmark, our main theorem establishes that this block-by-block guarantee translates directly to the global offline setting.

\begin{itheorem}[Informal, see \Cref{thm:myopic-upper-bound}]
	Every algorithm that is myopically-$\lambda$-optimal with respect to a fractional or integral benchmark yields a $\frac{\lambda}{1+\lambda}$ approximation to the corresponding optimal global offline allocation.
\end{itheorem}

From this general theorem, we immediately obtain efficient online algorithms for two fundamental settings. 
First, in the purely fractional relaxation of the problem, optimizing each block myopically via linear programming yields an exact fractional solution ($\lambda=1$), directly resulting in an online $1/2$-approximation (\Cref{cor:fractional-upper-bound}). Second, when the number of resources $m$ is a constant, we can apply known polynomial-time approximation schemes~\cite{FC84} to the single-block knapsack problem. Plugging a myopic $(1-\epsilon)$-approximation into our framework immediately provides an online $(1/2-\epsilon)$-approximation (\Cref{cor:constant-dim-upper-bound}).

Even when the number of resources $m$ is not constant, efficient approximations remain possible by leveraging the practical setting where transaction sizes are small relative to the block capacity ($q_{\max} \ll 1$). By deterministically truncating the myopic fractional solution for each block, we obtain an online algorithm with an approximation ratio of $(1-mq_{\max})/2$ (\Cref{cor:det-rounding-upper-bound}). While this is likely sufficient for practical applications, theoretically the linear degradation with $m$ can be improved and  we develop a randomized rounding algorithm (\Cref{thm:rand-rounding-upper-bound}) that achieves a polynomial-time constant-factor approximation even when transaction sizes scale up to $q_{\max} = O(1/\log m)$.

As mentioned above, for the fully general case (an arbitrary number of resources and unrestricted transaction sizes), obtaining an efficient approximation algorithm is impossible assuming $P \ne NP$. To bypass this impossibility, we can utilize \textbf{resource augmentation}. This relaxation is highly justified in the blockchain context: block capacities are not rigid physical boundaries, but rather flexible design compromises. Following Babaioff and Nisan~\cite{BN25}, we allow the online algorithm to temporarily exceed the strict block limit $B_j$. However, motivated by the actual mechanics of Ethereum's EIP-1559, this augmentation is tightly constrained: while the \emph{maximum} block size may increase, the algorithm guarantees that the \emph{average} block size remains strictly bounded by $B_j$ (a property termed \enquote{slackness}). The exact definitions and our resulting approximation theorem for this general setting appear in \Cref{app:general-setting}.

\subsubsection{Lower Bounds}
For the important practical cases of a small number of resources or small transaction sizes, our algorithmic framework achieves an approximation ratio of nearly $1/2$. A natural question is whether one can design online algorithms with significantly better guarantees. We provide two lower bounds (\Cref{thm:general-lower-bound,thm:myopic-lower-bound}) demonstrating the difficulty of improving this ratio, even in the highly favorable setting of fully patient users, arbitrarily small transaction sizes, and permitted resource augmentations.

Our first lower bound resolves an open question raised by Babaioff and Nisan~\cite{BN25}. While they proved that a constant fraction of welfare must be lost for $m \ge 3$ resources, we show that this impossibility result extends even to the minimal multidimensional case of $m=2$.

\begin{itheorem}[Informal, see \Cref{thm:general-lower-bound}]
    Every online algorithm for the block packing problem with $m \ge 2$ resources can achieve at most a $7/8 + o(1)$ fraction of the optimal social welfare even with fully patient users, arbitrarily small transaction sizes and resource augmentations as in \cite{BN25}.
\end{itheorem}

While no algorithm can exceed the $7/8$ bound, our second lower bound shows that merely extending our myopic techniques will not suffice to surpass the $1/2$ approximation ratio. To formalize this, we define a natural property of \enquote{reasonable} scheduling algorithms. We say that transaction $i_1$ \emph{strictly dominates} transaction $i_2$ if it provides strictly more value while consuming at most the same capacity across every resource: $v_{i_1} > v_{i_2}$ and $w_{i_1 j} \le w_{i_2 j}$ for all $j \in \{1, \dots, m\}$.

\begin{definition}
    An online allocation algorithm is said to be \emph{myopically reasonable} if it never schedules a dominated transaction $i_2$ while its dominating transaction $i_1$ remains available and unscheduled.
\end{definition}

While this property seems like a natural baseline for any sensible heuristic, our next result shows that it strictly limits the achievable welfare.

\begin{itheorem}[Informal, see \Cref{thm:myopic-lower-bound}]
    Every myopically reasonable algorithm for the block packing problem with $m \ge 2$ resources can achieve at most a $\frac{m}{2m-1} + o(1)$ fraction of the optimal social welfare (which approaches $1/2$ for large $m$) even with fully patient users,
    arbitrarily small transaction sizes and resource augmentations as in \cite{BN25}.
\end{itheorem}

This demonstrates that any algorithm aiming to break the $1/2$ barrier must occasionally make highly counter-intuitive, non-myopic choices.

The main algorithmic question left open is determining whether $1/2$ is indeed the best possible approximation ratio for the basic case of small $m$, small transaction sizes, and fully patient transactions. For the fully general case, we have established that some resource augmentation is necessary, but achieving bounds with weaker augmentations remains an interesting direction for future work.

\subsection{EIP-1559} \label{sub:eip-1559}
The transaction fee mechanism used by the Ethereum blockchain is termed EIP-1559~\cite{1559}. This mechanism combines several ingredients to address both algorithmic and strategic considerations. Roughly, it includes the following core elements: (1) increasing the maximum block size by a factor of $c=2$, while ensuring that the average block size is maintained; (2) determining a fixed formula for the \enquote{base fee} of gas, where the price increases following over-demand and decreases following under-demand; (3) burning this base fee rather than awarding it to the block builder;\footnote{While in Ethereum there can be a separation between the builder and the proposer, we use the term \emph{block builder} to refer to a single entity that encompasses both.} and (4) adding a \enquote{tip} mechanism that functions as a first-price-like auction to incentivize the builder.

Starting with~\cite{R21}, a large body of literature has studied the strategic properties of this mechanism (e.g., \cite{CS23, GTW24}), but its welfare guarantees have received considerably less attention~\cite{BN25, FMPS21}. In this section, we investigate the social welfare achieved by multidimensional variants of EIP-1559. Since EIP-1559 (like any major blockchain fee mechanism) does not provide a way for transactions to bid based on time sensitivity, we restrict our analysis in this section to fully patient bidders.

\subsubsection{The Mechanics of Multidimensional EIP-1559}

While originally defined for a single dimension, the adoption of EIP-4844 in 2024~\cite{4844} effectively transformed Ethereum into a multidimensional environment. To analyze this complex system, we abstract away the strategic tip component and focus strictly on the algorithmic interplay between two distinct entities: the \emph{protocol} (which dictates rules and prices) and the \emph{block builder} (which makes the actual scheduling decisions). 

Formally, the multidimensional EIP-1559 framework operates under the following rules:\footnote{While EIP-1559 allows users to specify maximum prices, adding this bidding parameter does not affect our welfare bounds. Additionally, actual implementations use a first-order Taylor approximation $1+\eta(B_j^t/B_j - 1)$ for the exponential update. Currently, the target block sizes are $30$M gas and $14$ blobs, with maximums of $60$M gas ($c=2$) and $21$ blobs ($c=1.5$), using a step size of $\eta=1/8$.}

\begin{itemize}
    \item \textbf{Target and Maximum Capacities:} For each resource $j \in \{1, \dots, m\}$, the protocol defines a target capacity $B_j$ and a capacity multiplier $c_j \ge 1$. The hard upper limit for resource $j$ in any single block is $c_j B_j$.
    \item \textbf{Base Prices and Transaction Validity:} The protocol maintains a base price $p_j^t$ for each resource at time $t$ (initialized to $p_1$, with a hard floor of $p_{\min}$). An unscheduled transaction $i$ is considered \emph{valid} for block $t$ only if its value covers its total resource cost at current prices: $v_i \ge \sum_{j=1}^m w_{ij} p_j^t$.
    \item \textbf{The Builder's Discretion:} The protocol delegates block construction to a block builder. The builder may select any feasible subset of the currently valid transactions, provided the total weight respects the maximum capacity $c_j B_j$ for all resources. 
    \item \textbf{The Price Update Rule:} After block $t$ is finalized, the protocol updates the price of each resource $j$ based on its actual utilization $B_j^t := \sum_{i} w_{ij} x_i^t$. The price is adjusted exponentially using a global parameter $\eta > 0$:
    \[
        p_j^{t + 1} = \max\left\{p_j^t \cdot e^{\eta(B_j^t/B_j - 1)}, p_{\min}\right\}
    \]
\end{itemize}

A crucial consequence of this division of labor is that the protocol's pricing mechanism alone does not dictate the allocation; the builder's strategic choices determine the final welfare. \enquote{Usually}, the block builder can easily schedule all valid transactions. However, \enquote{occasionally} (in Ethereum, for about $2\%$ of blocks), the maximum block size is insufficient to hold all valid transactions, forcing the builder to choose a subset. As we will show, in the multidimensional setting, these choices are paramount. 

Despite allowing blocks to scale up to $c_j B_j$, a significant algorithmic property of this price update rule is that it naturally regulates the average block size. Adapting the formalization of Babaioff and Nisan~\cite{BN25} to the multidimensional case, we define this property as follows:

\begin{definition} \label{dfn:slackness}
	An allocation $\{x^t_i\}_{i,t}$ is said to have an average block size of at most $\{B_j\}_{j=1}^m$ with slackness $\Delta$ if, for any interval of consecutive blocks $[T_1,T_2]$, the total allocation of each resource $j$ satisfies $\sum_{t=T_1}^{T_2} \sum_i x^t_i w_{ij} \le (T_2 - T_1 + 1 + \Delta) \cdot B_j$.
\end{definition}

As in the single-dimensional case~\cite{BN25}, we observe that the allocations produced by the EIP-1559 protocol, regardless of the block builder's behavior, maintain an average block size of at most $B_j$. Crucially, the slackness is independent of the horizon $T$; it is strictly bounded by a constant determined by the protocol parameters: $\Delta=\frac{1}{\eta}\ln\left(\frac{V_{\max}}{p_{\min}B_j}\right) + c_j - 1$, where $V_{\max}$ is the maximum social welfare obtainable in a single block (i.e., $\sum_i v_i x_i^t \le V_{\max}$). We formally prove this feasibility in \Cref{prop:1559-feasibility}.

\subsubsection{EIP-1559 with Myopically Optimizing Block Builders}

Our main positive result demonstrates that if the block builder acts as a \emph{myopically-$\lambda$-optimal} algorithm (as introduced in \Cref{sub:online-block-packing}), the combination of EIP-1559 and this builder successfully approximates the global offline optimal allocation. At 
any time $t$, the builder is restricted to scheduling only \emph{valid} transactions (those willing to pay the current protocol prices) and we only require the builder to be \emph{myopically-$\lambda$-optimal} relative to an integral benchmark that considers \emph{only this restricted valid set.}

This algorithmic abstraction assumes that the builder's strategic optimization of tips aligns with maximizing the true underlying values of the transactions. While we explicitly abstract away the auction mechanics, such alignment is a standard property of equilibrium behavior in these fee markets. Under these conditions, the protocol yields a strong global guarantee:

\begin{restatable}{theorem}{EIPUpperBound} \label{thm:1559-upper-bound}
    For every block building algorithm BUILD that is myopically-$\lambda$-optimal, input transactions sequence $\mathcal C$ and horizon $T$:
    \[
    SW_{[1:T]}(x) \ge \frac{\lambda}{1+\lambda+\lambda e^{\eta(c-1)}} \cdot \left(SW_{[1:T]}(y) - \sum_{j=1}^mB_j(p_1\tau + p_{\min}T)\right)
    \]
    where $c:=\max_j c_j$, $\tau := 1 + \ln(p_1/p_{\min})/\eta$, $x$ is the output of the EIP-1559 algorithm that uses BUILD, and $y$ is the optimal (offline) allocation for $T$ blocks with resource capacities $\{B_j\}_{j=1}^m$. 
\end{restatable}

Notice that this welfare guarantee
can be explained by three distinct sources of efficiency loss:
\begin{enumerate}
    \item \textbf{Myopic Gap ($\boldsymbol{\lambda}$):}  This factor captures the efficiency loss inherent in myopically optimizing a single block. Recall that any myopically reasonable builder is bounded by an approximation ratio of $m/(2m-1)$ (which approaches $1/2$ for large $m$). In comparison, under standard Ethereum parameters ($\lambda=1, c=2, \eta=1/8$), our bound yields a ratio of approximately $0.32$.
    \item \textbf{Floor Constraint ($\boldsymbol{p_{\min}T}$):} This additive term accounts for the potential rejection of valid transactions valued slightly below the protocol's minimum price $p_{\min}$. Since $p_{\min}$ is a configurable parameter, it can be set sufficiently low (e.g., $0.01\%$ of typical prices) to minimize this impact.
    \item \textbf{Price Discovery ($\boldsymbol{p_1 \tau}$):} If the initial price $p_1$ is significantly higher than transaction valuations, the protocol may leave blocks empty while prices decay, whereas an optimal offline allocator would utilize them immediately. Importantly, this cost is constant and becomes negligible as the horizon $T$ grows.
\end{enumerate}

Conceptually, this result complements recent work by Angeris et al.~\cite{ADM25}, who also show that multidimensional variants of EIP-1559 approximate the optimal social welfare. Crucially, however, our benchmark is the true offline optimum, whereas the benchmark in~\cite{ADM25} is the significantly weaker welfare obtained by static, fixed prices.\footnote{Our lower bound in the subsequent section highlights the importance of this distinction, demonstrating that the best fixed-price welfare can be worse than the true offline optimum by a factor of $m/c$.}

\subsubsection{Lower Bounds for Multidimensional EIP-1559}

Because the EIP-1559 algorithm regulates resource prices in a natural, market-like fashion, one might hope that the system's efficiency does not strictly depend on the exact block building algorithm. This would seem natural given the empirical rarity of blocks where the maximum capacity is insufficient to hold all valid transactions. Indeed, Babaioff and Nisan~\cite{BN25} proved that in the single-dimensional case, a social-welfare-efficient outcome is obtained as long as the block builder does not leave \enquote{money on the table}, that is, as long as it schedules \emph{any} maximal (by inclusion) subset of valid transactions.

Our main negative result demonstrates that this guarantee collapses in the multidimensional setting. To illustrate this, we analyze the most natural heuristic: the \emph{greedy block builder}. This builder prioritizes valid transactions in decreasing order of their value-to-cost ratio, $v_i/(\sum_j w_{ij}p_j^t)$, scheduling them sequentially as long as resource capacities permit.\footnote{For the single-dimensional case, this heuristic of ordering by $v_i/w_i$ yields the classical LP-relaxation. In actual strategic implementations, the true value $v_i$ is private to the transaction owner, who typically uses the tip as a proxy; in the best case, this results in the identical ordering.}

Crucially, because it iterates through all valid transactions, the greedy builder inherently produces a maximal-by-inclusion allocation. Nevertheless, it fails to provide any meaningful approximation in the multidimensional environment:

\begin{itheorem}[Informal, see \Cref{thm:1559-lower-bound}]
EIP-1559 paired with the greedy block builder algorithm cannot approximate the optimal social welfare to within a factor better than $c/m+o(1)$, where $c=\max_j c_j$. This lower bound holds even for fully patient bidders with arbitrarily small transaction sizes, and even against a fixed stochastic distribution of transactions.
\end{itheorem}

\subsubsection{Summary of Results and Paper Organization}

To synthesize our theoretical and practical contributions, \Cref{tab:summary-results} provides a comprehensive overview of our established welfare guarantees for both the pure algorithmic block packing problem and the EIP-1559 mechanism.

\begin{table}[htbp] 
\caption{Summary of the results}
\label{tab:summary-results}
\centering
\renewcommand{\arraystretch}{1.2}
\small
\begin{tabular}{@{} l l l c @{}} 
\toprule
\textbf{Setting} & \textbf{Positive (UB)} & \textbf{Negative (LB)} & \textbf{Ref.} \\ 
\midrule

% --- PART I ---
\multicolumn{4}{@{}l}{\textit{Part I: Pure Algorithmic Framework}} \\
Myopically-$\lambda$-Optimal Algorithm & $\lambda/(1+\lambda)$ & & \Cref{thm:myopic-upper-bound} \\
Fractional Block Packing & $1/2$ & & \Cref{cor:fractional-upper-bound} \\
Constant Dimensions (PTAS) & $1/2 - \epsilon$ & & \Cref{cor:constant-dim-upper-bound} \\
Small Transactions & $1/2-mq_{\max}$ & & \Cref{cor:det-rounding-upper-bound} \\
Bounded Size Transactions $q_{\max} = O(1/\log m)$ & $\Omega(1)$ & & \Cref{thm:rand-rounding-upper-bound} \\
General Multidimensional Setting ($m \ge 2$) & & $7/8 + o(1)$ & \Cref{thm:general-lower-bound} \\
Myopically Reasonable Algorithm & & $\frac{m}{2m-1} + o(1)$ & \Cref{thm:myopic-lower-bound} \\
\midrule

% --- PART II ---
\multicolumn{4}{@{}l}{\textit{Part II: Multidimensional EIP-1559}} \\
Myopically-$\lambda$-Optimal Builder & $\approx \frac{\lambda}{1+\lambda+\lambda e^{\eta(c-1)}}$ & & \Cref{thm:1559-upper-bound} \\
Greedy Builder & & $c/m + o(1)$ & \Cref{thm:1559-lower-bound} \\

\bottomrule
\end{tabular}
\end{table}

\paragraph{Organization of the Paper.} 
The remainder of this paper is organized as follows. \Cref{sec:algorithms} focuses on the pure algorithmic block packing problem, where we introduce our myopic reduction framework and establish constant-factor approximation algorithms for both the fractional setting and the integral setting with small transaction sizes. Next, \Cref{sec:lower-bounds} establishes the fundamental limitations of online block packing, proving the strict impossibility of near-optimal approximations even for two resources, alongside bounds constraining any myopically reasonable scheduling algorithm. In \Cref{sec:1559}, we transition to evaluating the multidimensional EIP-1559 mechanism in practice, formalizing the positive global welfare guarantees when the protocol is paired with a myopically optimal block builder, and demonstrating the structural vulnerabilities of the natural greedy heuristic. Finally, our general-setting resource augmentation framework is provided in \Cref{app:general-setting}.

\section{Online Block Packing Algorithms} \label{sec:algorithms}

The optimization problem faced by our algorithm at every step, i.e., for every block,
is exactly the classical multidimensional knapsack problem~\cite{KPP04}. In this classic optimization problem, a set of items, each with a value and a vector of weights, must be packed into a knapsack with multiple resource constraints so as to maximize the total value of selected items without exceeding any resource capacity. In our context, each block corresponds to a knapsack with $m$ resource constraints $\{B_j\}_{j=1}^m$, and each transaction is an item characterized by its value $v_i^t=v_i(1-\rho_i)^{t-a_i}$ at time $t$ and an $m$-dimensional resource usage vector $w_i\in\mathbb{R}^m_+$.

The main difference is that in our online context, we do not aim to optimize for each block separately but rather for the whole sequence of blocks. Even exact optimization for each block (which is not computationally feasible) does not ensure
global social welfare optimization. We start this
section by showing that,
just like in the fractional case,
such block-by-block approximate optimization does imply good approximation overall. We then
continue by presenting approximation algorithms for the single block multidimensional knapsack
problem, for the special case of small transactions.

\subsection{Repeated Myopic Solution of Multidimensional Knapsack}

This subsection studies a myopic approach: at each time \(t\) the remaining transactions form a multidimensional knapsack instance, which we solve by invoking an approximation oracle.

\begin{algorithm}[H]
    \caption{Block Packing via Multidimensional Knapsack (MK) Oracle}
    \label{alg:myopic-via-oracle}
    
    \SetKwInOut{Input}{Input}
    \SetKwInOut{Uses}{Uses}
    \SetKwInOut{Output}{Output}

    \Input{Transaction set $\{(a_i, v_i, \rho_i, \{w_{ij}\}_{j=1}^m)\}_{i \in \mathcal{C}}$; Block capacities $\{B_j\}_{j=1}^m$}
    \Uses{Multidimensional Knapsack Oracle ALG}
    \Output{Allocation $\{x_i^t\}_{i \in \mathcal{C},\, t \in \mathbb{N}}$ where $x_i^t \in \{0,1\}$}

    \BlankLine
    \For{each time step $t = 1, 2, \dots$}{
        Define $U_t$ as the set of available transactions that have arrived by time $t$ and have not been included in any previous block $s < t$:
        \[
            U_t = \{i \in \mathcal{C} \mid a_i \le t \text{ and } \forall s < t, x_i^s = 0\}
        \]
        
        Compute $\{x^t_i\}_{i\in\mathcal C}$ by calling ALG with resource capacities $\{B_j\}_{j=1}^m$ and available transactions $U_t$, where each transaction $i$ has a discounted value $v_i^t = v_i (1-\rho_i)^{t-a_i}$\;
    }
\end{algorithm}

The main theorem in this section states that if our multidimensional knapsack oracle produces good approximations at every
stage, then for every horizon we also get a good approximation overall.

Let $OPT^t$ denote the value of the optimal integral
solution for the multidimensional knapsack problem at time $t$ and $ALG^t$ denote the value of the solution produced
by the multidimensional knapsack oracle at time $t$, i.e., $ALG^t = \sum_i x_i^t v_i^t$. 

\begin{theorem} \label{thm:myopic-upper-bound}
	If for every $t$ we have that
	$ALG^t \ge \lambda \cdot OPT^t$ then for any integral allocation
	$y=\{y_i^t\}$ and for every horizon $T$ we have that $SW_{[1:T]}(x)\ge \frac{\lambda}{1+\lambda}\cdot SW_{[1:T]}(y)$.

	Similarly, if for every $t$ we have that $ALG^t \ge \lambda \cdot OPT^{*t}$, where $OPT^{*t}$ denotes the value of the
	optimal \emph{fractional} solution for time $t$, then we also have that for any \emph{fractional} solution $y^* = \{y^{*t}_i\}$ it holds that $SW_{[1:T]}(x)\ge \frac{\lambda}{1+\lambda}\cdot SW_{[1:T]}(y^*)$.
\end{theorem}

The theorem also holds when the output of the algorithm is a \emph{fractional} allocation, i.e., $x_i^t\in[0,1]$.
This allows us to reduce the global block packing problem to optimizing each block independently. In particular, if we can find a $(1-\epsilon)$-approximation to the myopic fractional solution for each block, this yields a $\left(\frac{1}{2} - \epsilon\right)$-approximation to the optimal offline fractional allocation.
The proof relies on the following lemma.

\begin{lemma} \label{lem:myopic-helper}
	Let $x=\{x_i^t\}_{i,t},y=\{y_i^t\}_{i,t}$ be two arbitrary, possibly fractional, allocations. Denote by $X_i^t=\sum_{\tau=1}^tx_i^{\tau}$ the cumulative allocation of transaction $i$ up to time $t$ made by $x$. Then for every transaction $i$ and horizon $T$ it holds:
	\[
		\sum_{t=1}^T x_i^tv_i^t \ge \sum_{t=1}^T X_i^t y_i^t v_i^t
	\]
\end{lemma}

Intuitively, $X_i^t$ quantifies the amount of transaction $i$ allocated by $x$ until time $t$ and the term $X_i^ty_i^t$ denotes the portion of transaction $i$ that is allocated by $y$ in block $t$ and by $x$ until block $t$. The lemma states that the contribution to the social welfare from the part of $y$'s allocation that was scheduled after $x$'s allocation is no greater than the social welfare obtained by $x$.

\begin{proof}
	Let $T \ge 1$. For any transaction $i$, we have:
	\[
		\sum_{t=1}^T X_i^t y_i^t v_i^t = \sum_{t=1}^T y_i^t \sum_{\tau=1}^t x_i^\tau v_i^t
		\le \sum_{t=1}^T y_i^t \sum_{\tau=1}^t x_i^\tau v_i^\tau
		\le \sum_{t=1}^T y_i^t \sum_{\tau=1}^T x_i^\tau v_i^\tau
		\le \sum_{\tau=1}^T x_i^\tau v_i^\tau,
	\]
	where we used that $v_i^t$ is non-increasing in $t$ and $\sum_t y_i^t \le 1$.
\end{proof}

Using \Cref{lem:myopic-helper}, we now proceed to the proof of \Cref{thm:myopic-upper-bound}.

\begin{proof}[Proof of \Cref{thm:myopic-upper-bound}]
	We will prove the theorem for approximating integral solutions; the proof for the fractional case is identical.
	Fix a horizon $T$.
	Denote by $x=\{x_i^t\}_{i,t}$ the output of the algorithm,
	$z^t=\{z^t_i\}_i$ the optimal integral \enquote{myopic} allocations over transactions in $U_t$ for every time $t$,
	and $y=\{y_i^t\}_{i,t}$ any integral allocation.
	Assume $\sum_i x_i^t v_i^t \ge\lambda\cdot \sum_i z_i^t v_i^t$ for every $t$.

	For each $t$, define $\tilde y_i^t := (1 - X_i^{t-1}) y_i^t$, for each transaction $i$, where $X_i^{t-1} = \sum_{\tau < t} x_i^\tau$.
    We show that $\tilde y^t$ is a feasible solution for the multidimensional knapsack problem at time $t$.
    Since $\tilde y_i^t\le 1-X_i^{t-1}$, the total allocation of each individual transaction does not exceed $1$. From the feasibility of $y^t$ and $\tilde y_i^t \le y_i^t$, the vector $\tilde y^t = \{\tilde y_i^t\}_i$ is a feasible solution over the available transactions $U_t$ at time $t$.

	By the optimality of $z^t$ and the $\lambda$-approximation guarantee of $x^t$, we have:
	\[
		\sum_i x_i^t v_i^t \ge \lambda \cdot \sum_i z_i^t v_i^t \ge \lambda \sum_i \tilde y_i^t v_i^t = \lambda \sum_i (1 - X_i^{t-1}) y_i^t v_i^t.
	\]
	Summing over $t$ gives:
	\[
		SW_{[1:T]}(x) = \sum_{t=1}^T \sum_i x_i^t v_i^t \ge \lambda \sum_{t=1}^T \sum_i (1 - X_i^{t-1}) y_i^t v_i^t.
	\]
	On the other hand, from \Cref{lem:myopic-helper} we have:
	\[
		SW_{[1:T]}(x)=\sum_{t=1}^T\sum_ix_i^tv_i^t \ge \sum_{t=1}^T\sum_iX_i^ty_i^tv_i^t
	\]
	Combining, we get:
	\[
		\left(\frac{1}{\lambda} + 1\right)SW_{[1:T]}(x) \ge
		\sum_{t=1}^T \sum_i (1 - X_i^{t-1}) y_i^t v_i^t + \sum_{t=1}^T\sum_iX_i^ty_i^tv_i^t,
	\]
	and because $(1 - X_i^{t-1}) + X_i^t = 1 + x_i^t \ge 1$:
	\[
		SW_{[1:T]}(x) \ge \frac{\lambda}{1 + \lambda} SW_{[1:T]}(y),
	\]
	as claimed.
\end{proof}

We can now use known multidimensional knapsack algorithms and apply \Cref{thm:myopic-upper-bound} for approximating the online block packing problem.  

\begin{corollary} \label{cor:fractional-upper-bound}
    The myopic fractional algorithm (\Cref{alg:myopic-via-oracle}) that solves each multidimensional fractional knapsack using LP yields a $1/2$-approximation for the online fractional block packing problem.
\end{corollary}

\begin{corollary} \label{cor:constant-dim-upper-bound}
    For constant $m$, for every $\epsilon > 0$ there exists an online algorithm for the integral block packing problem that obtains a $(1/2-\epsilon)$-approximation for the optimal offline fractional solution.
\end{corollary}

\begin{proof}
    For constant $m$ there are known PTAS algorithms for the multidimensional knapsack problem (\cite{FC84}), obtaining a $(1-\epsilon)$-approximation for any $\epsilon > 0$.
    Applying \Cref{alg:myopic-via-oracle} yields a $(1/2-\epsilon)$-approximation for the online block packing problem.
\end{proof}

\subsection{MK Approximation Algorithms for the Case of Small Transactions} \label{sub:det-rounding}

As previously mentioned, approximating the multidimensional block packing problem in the general setting is NP-hard.
We give a very simple rounding algorithm for the case of
\enquote{tiny} transactions, $q_{\max}:=\max_{i,j} \frac{w_{ij}}{B_j} < \frac{1}{m}$.

\begin{algorithm}[H]
    \caption{LP Rounding for Multidimensional Knapsack}
    \label{alg:det-rounding}
    
    \SetKwInOut{Input}{Input}
    \SetKwInOut{Output}{Output}

    \Input{Item set $\{(v_i, \{w_{ij}\}_{j=1}^m)\}_{i \in \mathcal{C}}$; Resource constraints $\{B_j\}_{j=1}^m$}
    \Output{Integral allocation $\{x_i\}_{i\in\mathcal C}$}

    \BlankLine
    Find the myopic fractional allocation, a \emph{basic} solution $\{x_i^*\}_{i\in\mathcal C}$ for the following linear program:
    \[
        \text{\textbf{maximize}} \quad \sum_i x_i^* \cdot v_i \quad
        \text{\textbf{subject to}} \quad
        \begin{cases}
            0 \le x_i^* \le 1 & \text{for all } i, \\
            \sum_{i\in\mathcal C} w_{ij} \cdot x_i^* \le B_j & \text{for all } j.
        \end{cases}
    \]
    
    Return the rounded integral solution:
    \[
        x_i \;=\;
        \begin{cases}
            1 & \text{if } x_i^*=1, \\
            0 & \text{otherwise},
        \end{cases}
        \qquad\text{for all } i.
    \]
\end{algorithm}

\begin{lemma} \label{lem:det-rounding}
	For every $\delta>0$, if $q_{\max}:=\max_{i,j}\frac{w_{ij}}{B_j}\le\frac{\delta}{m}$ then LP rounding (\Cref{alg:det-rounding}) obtains a $(1-\delta)$-approximation for the optimal fractional solution:
	\[
		\sum_{i\in\mathcal C}x_iv_i \ge (1-\delta)\sum_{i\in\mathcal C}x_i^*v_i
	\]
\end{lemma}

As a direct corollary from this lemma, we get the following result:
\begin{corollary} \label{cor:det-rounding-upper-bound}
    For every $\delta>0$, if $q_{\max}\le \delta/m$ then solving each block myopically (\Cref{alg:myopic-via-oracle}) using LP rounding (\Cref{alg:det-rounding}) as the multidimensional knapsack oracle yields a $\frac{1-\delta}{2}$-approximation algorithm for the online block packing problem.
\end{corollary}

The proof of \Cref{lem:det-rounding} relies on another lemma that bounds the value of each fractionally allocated transaction. Denote the values of the fractional and rounded solutions by:
\[
	V^{*}:=\sum_{i\in\mathcal C}v_i x_i^{*},
	\qquad
	V:=\sum_{i\in\mathcal C}v_i x_i.
\]

\begin{lemma} \label{lem:det-rounding-helper}
	Let $k$ be some fractionally allocated transaction, i.e., $0<x_k^*<1$. Then the value of transaction $k$ is bounded by $v_k \le q_{\max}V^*$.
\end{lemma}

\begin{proof}
	Intuitively, since \(x^*\) is an optimal fractional solution with total value \(V^*\), any feasible perturbation cannot increase the objective.
	Shrink every allocation by a small factor \((1-\varepsilon)\) and use the freed capacity to add \(\varepsilon/q_{\max}\) units of the still–fractional transaction \(k\).
	All constraints remain satisfied, and the objective changes by \(\varepsilon\bigl(v_k/q_{\max}-V^*\bigr)\), which must be non-positive.
	Formally, define the following fractional allocation:
	\[
		y_i :=
		\begin{cases}
			(1-\varepsilon) x_i^{*}                               & i\ne k, \\
			(1-\varepsilon) x_k^{*}+\dfrac{\varepsilon}{q_{\max}} & i=k,
		\end{cases}
		\quad \text{for all }i.
	\]

	For sufficiently small \(\varepsilon\), this solution is feasible:
	\[
		\sum_{i}w_{ij}y_i =
		(1-\varepsilon)\sum_{i}w_{ij}x_i^{*} + \frac{\varepsilon}{q_{\max}}\,w_{kj}
		\le
		(1-\varepsilon) B_j+\varepsilon B_j=B_j,
	\]
	and for $\varepsilon \le q_{\max}(1-x_k^*)$ we have
	$
		y_k \le x_{k}^{*}+\frac{\varepsilon}{q_{\max}}\le1
	$.

	The optimality of $x^*$ implies
	\(\sum_{i}v_i x_i^{*}\ge \sum_{i}v_i y_i\), i.e.,
	\[
		\sum_{i}v_{i}\left(x_{i}^{*}-y_{i}\right) =
		\varepsilon\left(\sum_{i}v_{i}x_{i}^{*}-\frac{v_{k}}{q_{\max}}\right) =
		\varepsilon\left(V^{*}-\frac{v_{k}}{q_{\max}}\right) \ge 0,
	\]
	resulting in $v_k\le q_{\max}V^*$.
\end{proof}

\begin{proof} [Proof of~\Cref{lem:det-rounding}]
	Denote the set of fractionally allocated transactions by
	\[
		F=\{i\in\mathcal{C}\mid 0< x_i^{*}<1\}
	\]
	Because the LP has \(m\) constraints, any basic optimal
	solution satisfies \(|F|\le m\). The rounding discards only the items in \(F\), and using~\Cref{lem:det-rounding-helper} we get
	\[
		V^{*}-V
		=\sum_{i\in F}v_i x_i^{*}
		\le\sum_{i\in F}v_i
		\le |F|\;q_{\max}V^{*}
		\le m q_{\max}V^{*}
		\le\delta V^{*}.
	\]
	Hence \(V\ge(1-\delta)V^{*}\), proving the claimed
	\((1-\delta)\)-approximation.
\end{proof}

\subsection{Randomized Rounding for 'Not Too Large' Transactions} \label{sub:rand-rounding}

In \Cref{sub:det-rounding}, we showed that if transaction sizes are small enough, namely $q_{\max}:=\max_{i,j}\frac{w_{ij}}{B_j} < \frac{1}{m}$, then we can obtain a constant approximation for the online block packing problem using myopic block building that uses LP rounding (\Cref{cor:det-rounding-upper-bound}). We show that more advanced randomized rounding techniques can approximate the optimal social welfare even when transactions are significantly larger, namely $q_{\max} = O\left(\frac{1}{\log m}\right)$.

\begin{theorem} \label{thm:rand-rounding-upper-bound}
    For every $\delta>0$, if $q_{\max}\le c \cdot \frac{\delta^2}{\log(m/\delta)}$ for a small enough constant $c>0$, then solving each block myopically (\Cref{alg:myopic-via-oracle}) using the randomized rounding algorithm (\Cref{alg:randomized-rounding}) as the multidimensional knapsack oracle yields a $\frac{1-\delta}{2}$-approximation algorithm for the online block packing problem.
\end{theorem}

\begin{algorithm}[H]
    \caption{Randomized Rounding for Multidimensional Knapsack}
    \label{alg:randomized-rounding}
    
    \SetKwInOut{Input}{Input}
    \SetKwInOut{Output}{Output}

    \Input{Item set $\{(v_i, \{w_{ij}\}_{j=1}^m)\}_{i \in \mathcal{C}}$; Resource constraints $\{B_j\}_{j=1}^m$; Parameter $\delta \in (0,1)$}
    \Output{Integral allocation $\{x_i\}_{i\in\mathcal{C}}$}
    Find the myopic fractional allocation, a \emph{basic} solution $\{x_i^*\}_{i\in\mathcal{C}}$ for the following linear program:
    \[
        \textbf{maximize} \quad \sum_{i\in\mathcal{C}} x_i^* \cdot v_i \quad
        \textbf{subject to} \quad
        \begin{cases}
            0 \le x_i^* \le 1 & \text{for all } i, \\
            \sum_{i\in\mathcal{C}} w_{ij} \cdot x_i^* \le B_j & \text{for all } j.
        \end{cases}
    \]
    \Repeat{both of the following conditions are met: \\
        \quad (1) $\sum_i w_{ij} x_i \le B_j$ for all $j$, \\
        \BlankLine
        \quad (2) $\sum_i v_i x_i \ge (1 - \delta) \sum_i v_i x_i^*$
    }{
        Independently sample each $x_i \sim \mathrm{Ber}\left((1-\frac{\delta}{2}) \cdot x_i^*\right)$
    }
    \BlankLine
    \Return $\{x_i\}_{i\in\mathcal{C}}$
\end{algorithm}

\begin{remark}
	Although the algorithm is randomized, it can likely be derandomized using the method of conditional expectations~\cite{R88}, but we have not formally confirmed this.
\end{remark}

\begin{lemma} \label{lem:rand-rounding-helper}
	Assume that \( q_{\max} := \max_{i,j} \frac{w_{ij}}{B_j} \le c \cdot \frac{\delta^2}{\log(m/\delta)} \) for a small enough constant $c>0$; then the randomized rounding algorithm (\Cref{alg:randomized-rounding}) terminates after an expected $O(m/\delta)$ iterations and returns a solution that is both feasible (i.e., satisfies all resource constraints) and achieves a $(1 - \delta)$-approximation for the optimal fractional solution.
\end{lemma}

\begin{proof} [Proof of \Cref{thm:rand-rounding-upper-bound}]
	This follows directly from \Cref{thm:myopic-upper-bound,lem:rand-rounding-helper} with the randomized rounding algorithm (\Cref{alg:randomized-rounding}) as the oracle.
\end{proof}

\Cref{lem:rand-rounding-helper} follows from the following two lemmas that state the appropriate tail bounds.

\begin{lemma} \label{lem:rand-rounding-size}
	For any fixed resource \( j \in [m] \) and accuracy parameter \( \delta \in (0,1) \), the probability that the resource constraint for $j$ is violated is bounded by:
	\[
		\mathbb{P}\left( \sum_i w_{ij} x_i > B_j \right) \le
		2\exp\left( -\frac{\delta^2}{8q_{\max}} \right).
	\]
\end{lemma}

\begin{lemma} \label{lem:rand-rounding-sw}
	For any \( \delta \in (0,1) \), the probability that the total value of the rounded solution is at least $(1 - \delta)$ times the fractional optimum is bounded by:
	\[
		\mathbb{P}\left( \sum_i v_i x_i \ge (1 - \delta) \sum_i v_i x_i^* \right) \ge \frac{\delta}{2(m + 1)}.
	\]
\end{lemma}

Given these two lemmas we can easily prove \Cref{lem:rand-rounding-helper}.

\begin{proof}[Proof of \Cref{lem:rand-rounding-helper}]
	In each iteration, the algorithm samples independent \( x_i \sim \mathrm{Ber}((1 - \tfrac{\delta}{2})x_i^*) \). Let us bound the probability that the sampled solution is both feasible and achieves value at least \( (1 - \delta) \sum_i v_i x_i^* \).

	By \Cref{lem:rand-rounding-size}, for each resource \( j \in [m] \),
	\[
		\mathbb{P}\left( \sum_i w_{ij} x_i > B_j \right) \le 2\exp\left( -\frac{\delta^2}{8q_{\max}} \right).
	\]
	By \Cref{lem:rand-rounding-sw},
	\[
		\mathbb{P}\left( \sum_i v_i x_i \ge (1 - \delta) \sum_i v_i x_i^* \right)
		\ge \frac{\delta}{2(m + 1)}.
	\]

	Using the union bound over all \( m \) constraints and the value condition:
	\[
		\mathbb{P}(\text{failure})
		\le 2m \cdot \exp\left( -\frac{\delta^2}{8q_{\max}} \right) +
		1 - \frac{\delta}{2(m + 1)}.
	\]

	Now assume \( q_{\max} \le \frac{\delta^2}{16 \log(5m/\delta)} \). Then:
	\[
		\exp\left( -\frac{\delta^2}{8q_{\max}} \right) \le \exp\left( -2 \log(5m/\delta) \right) = \left( \frac{\delta}{5m} \right)^2,
	\]
	therefore,
	\[
		\mathbb{P}(\text{success})
		\ge \frac{\delta}{2(m + 1)} - 2m \cdot \left( \frac{\delta}{5m} \right)^2
		\ge \frac{\delta}{4m} - \frac{2}{25}\frac{\delta}{m} > \frac{\delta}{10m}.
	\]

	Thus, the algorithm halts with probability 1 and has an expected number of iterations \( O(m/\delta) \).
\end{proof}

We now provide the technical proofs of
\Cref{lem:rand-rounding-size} and \Cref{lem:rand-rounding-sw} which use standard randomized rounding techniques. We will use Bernstein's inequality:

\begin{ktheorem}[Bernstein's Inequality] \cite[Thm.~2.9.5, p.~49]{Vershynin26}
	Let \( X_1, X_2, \dots, X_n \) be independent, mean-zero random variables such that \( |X_i| \le M \) for all \( i \), and define the total variance \( \sigma^2 := \sum_{i=1}^n \mathbb{E}[X_i^2] \). Then for any \( t > 0 \),
	\[
		\mathbb{P}\left( \left|\sum_{i=1}^n X_i \right| \ge t \right) \le 2\exp\left( - \frac{t^2}{2\sigma^2 + \frac{2}{3} M t} \right).
	\]
\end{ktheorem}

\begin{proof} [Proof of \Cref{lem:rand-rounding-size}]
	Fix \( j \in [m] \), and define:
	\[
		Y_i := w_{ij}(x_i - (1 - \tfrac{\delta}{2})x_i^*).
	\]
	Then \(\{Y_i\}\) are independent, mean-zero random variables satisfying:
	\[
		|Y_i| \le w_{ij} \le B_j q_{\max}
		\quad \text{and} \quad
		\sum_i \mathrm{Var}(Y_i) \le \sum_i w_{ij}^2 \cdot (1 - \tfrac{\delta}{2}) x_i^* \le (1 - \tfrac{\delta}{2})B_j^2 q_{\max}.
	\]

	From the feasibility of the fractional solution, we have:
	\[
		\sum_i w_{ij} x_i = \sum_i Y_i + (1 - \tfrac{\delta}{2}) \sum_i w_{ij} x_i^* \le \sum_i Y_i + (1 - \tfrac{\delta}{2}) B_j,
	\]
	and therefore,
	\[
		\mathbb{P}\left( \sum_i w_{ij} x_i > B_j \right) \le
		\mathbb{P}\left( \left|\sum_i Y_i\right| \ge \tfrac{\delta}{2} B_j \right).
	\]

	Apply Bernstein's inequality with deviation $t=\frac{\delta}{2}B_j$, variance $\sigma^2 \le (1-\frac{\delta}{2})B_j^2q_{\max}$, and bound $M=B_jq_{\max}$:
	\[
		\mathbb{P}\left(\left| \sum_i Y_i \right|\ge \tfrac{\delta}{2} B_j \right)
		\le
		2\exp\left( - \frac{\tfrac{\delta^2}{4} B_j^2}{2 (1-\frac{\delta}{2})B_j^2q_{\max} + \tfrac{2}{3} \frac{\delta}{2}B_j^2q_{\max}} \right) \le 2\exp\left( - \frac{\delta^2}{8 q_{\max}} \right),
	\]
	using \( \delta \le 1 \).
\end{proof}

\begin{proof}[Proof of \Cref{lem:rand-rounding-sw}]
	Let \(V=\sum_i v_i x_i\) be the algorithm's (random) welfare and \(V^{*}=\sum_i v_i x^{*}_i\) the fractional optimum.
	Optimality of \(x^{*}\) implies \(v_i\le V^{*}\) for every item \(i\); otherwise, taking that single item would yield a higher value.

	Write \(F=\{i:0<x^{*}_i<1\}\). Because $x^*$ is a basic solution, we have that $|F| \le m$, thus
	\[
		V=\sum_{i\in F} v_i x_i+\sum_{i\notin F} v_i x_i
		\le |F|\max_i v_i + V^{*}\le (m+1)V^{*}.
	\]

	Define \(p:=\mathbb{P}(V\ge(1-\delta)V^{*})\). Therefore,
	\[
		\mathbb{E}[V]\le (1-p)\cdot(1-\delta)V^{*}+p\cdot (m+1)V^{*}
	\]
	and because the expected value is \(\mathbb{E}[V]=\left(1-\tfrac{\delta}{2}\right)V^{*}\),
	\[
		1-\frac{\delta}{2}\le (1-p)(1-\delta)+p(m+1)
		\quad\Longrightarrow\quad
		p\ge\frac{\delta}{2(m+1)},
	\]
	as claimed.
\end{proof}

\section{Lower Bounds} \label{sec:lower-bounds}

In order to strengthen our lower bounds, we will prove them even when the algorithm is allowed two resource augmentations as in \cite{BN25}.  The first is allowing the algorithm to occasionally exceed the maximum block size, as long as the average block size is not exceeded, see \Cref{dfn:slackness} for the formal definition of \enquote{average block size with slackness}. The second resource augmentation is to allow the algorithm to run for $o(T)$ extra steps.

\subsection{Any Online Algorithm Must Lose a Constant Fraction even for \texorpdfstring{$m=2$}{m=2}}

\begin{theorem} \label{thm:general-lower-bound}
	Let \( m \ge 2 \). Fix an online fractional scheduling algorithm which satisfies an average block size constraint \( \{B_j\}_{j=1}^m \) with slackness \( \Delta(T) = o(T) \). Then for every extension \( \Gamma(T) = o(T) \) and horizon $T$ there exists an input with fully patient bidders such that the output of the algorithm \( x = \{x^t_i\}_{i,t} \) loses a constant fraction of the social welfare, i.e.,
	\[
		\frac{SW_{[1:T+\Gamma]}(x)}{SW_{[1:T]}(y)} \le \frac{7}{8}+o(1),
	\]
	where \( y=\{y_i^t\}_{i,t} \) is the optimal integral allocation (for horizon $T$) with per-block limits \( \{B_j\}_{j=1}^m \).
\end{theorem}

\begin{proof}
	It suffices to analyze the case of two unit-capacity resources, $B_1=B_2=1$, as the proof extends to the general case by scaling the sizes of all input transactions.

	\paragraph{Transaction types.} The adversary will use the following transaction types:
	\[
		\begin{array}{c|c|c}
			\text{type} & \text{value }v & \text{demand }(w_1,w_2) \\ \hline
			A           & 2              & (1,0)                   \\
			B           & 1              & (1,1)                   \\
			C           & 1              & (0,1)                   \\
		\end{array}
	\]
	Intuitively, a block can schedule either one \(B\) transaction or one disjoint pair \((A,C)\).

	\paragraph{Intuition.}
	The adversary exploits the algorithm's decisions.
	Initially, it releases an equal supply of type-$A$ and type-$B$ transactions. If the algorithm schedules few $A$ transactions, the adversary reveals more $A$ transactions, allowing the optimal solution to achieve higher welfare by allocating only $A$ transactions. Conversely, if the algorithm schedules many $A$ transactions, the adversary introduces type-$C$ transactions. This forces the algorithm to choose between the remaining $B$ transactions and the new $C$ transactions. Meanwhile, the optimal solution can schedule all $B$ transactions in the first phase and $(A,C)$ pairs in the second phase, attaining maximal welfare.

	\paragraph{Phase~1 (first $k:=T/2$ blocks).}
	At time~1 the adversary supplies $k$ copies of each of $A,B$.
	Let
	\[
		a_1=\frac1k\sum_{t=1}^{k}\!\sum_{i\in A}x_i^t,
		\quad
		b_1=\frac1k\sum_{t=1}^{k}\!\sum_{i\in B}x_i^t
	\]
	denote the fractions of allocation of each transaction type in the first phase. Denote the \emph{normalized} slackness, extension and social welfare of the algorithm by
	\[
		\delta:=\frac{\Delta(T)}{k},
		\quad
		\gamma:=\frac{\Gamma(T)}{k},
		\quad
		\omega:=\frac{SW_{[1:T+\Gamma]}(x)}{k}.
	\]
	Feasibility during the first half gives
	\begin{equation}\label{eq-1}
		a_1 + b_1 \le 1 + \delta.
	\end{equation}

	\paragraph{Phase~2 (last $k+\Gamma$ blocks).}
	Similarly, denote by $a_2,b_2$ the fraction of allocation of each transaction type in the second phase.
	The adversary then chooses the additional input transactions based on the value of $a_1$, and we split our analysis into two distinct cases.

	\paragraph{Case~(I) $\boldsymbol{a_1\le\tfrac12}$.}
	In this case, the adversary reveals $k$ additional transactions of type $A$ arriving at time $k+1$.
	Resource feasibility in the second phase gives
	\begin{equation}\label{eq-2}
		a_2 + b_2 \le 1 + \gamma + \delta,
	\end{equation}
	and from transaction supply constraints,
	\begin{equation}\label{eq-3}
		a_1 + a_2 \le 2, \quad b_1 + b_2 \le 1.
	\end{equation}

	Using~\eqref{eq-1}--\eqref{eq-3}, we can bound the welfare:
	\begin{align*}
		\omega
		 & =2(a_1 + a_2) + b_1 + b_2 \\
		 & \le
		\left(a_{1}+b_{1}\right)+a_{1}+2\left(a_{2}+b_{2}\right) \\
		 & \le
		\left(1+\delta\right)+\frac{1}{2}+2\left(1+\delta+\gamma\right)
		=\frac{7}{2}+o\left(1\right),
	\end{align*}
	which yields the following bound:
	\[
		SW_{[1:T+\Gamma]}(x) = k\cdot\omega \le
		\frac{7}{4}T+o(T).
	\]

	The offline optimum takes a single $A$ transaction in every block and obtains a total value of $4k=2T$. Thus, the competitive ratio is bounded by
	\[
		\frac{SW_{[1:T+\Gamma]}(x)}{SW_{[1:T]}(y)}
		=\frac{7}{8}+o(1).
	\]
	\paragraph{Case~(II) $\boldsymbol{a_1>\tfrac12}$.}
	In this case, the adversary supplies $k$ additional transactions of type $C$ arriving at time $k+1$, and their fraction of allocation is denoted by $c$.

	Similarly, resource constraints give
	\begin{equation}\label{eq-4}
		b_{2}+a_{2} \le 1+\delta+\gamma,
		\quad
		b_{2}+c \le 1+\delta+\gamma,
	\end{equation}
	and for each transaction type we have
	\begin{equation} \label{eq-5}
		a_{1}+a_{2} \le 1,
		\quad
		b_{1}+b_{2} \le 1,
		\quad
		c \le 1.
	\end{equation}

	Again by~\eqref{eq-1}, \eqref{eq-4}, \eqref{eq-5} we have
	\begin{align*}
		\omega
		 & = 2\left(a_{1}+a_{2}\right)+b_{1}+b_{2}+c \\
		 & \le
		2\left(a_{1}+a_{2}\right)+\left(a_{1}+b_{1}\right)-a_{1}+\left(b_{2}+c\right) \\
		 & \le
		2+\left(1+\delta\right)-\frac{1}{2}+\left(1+\delta+\gamma\right)
		=\frac{7}{2}+o(1),
	\end{align*}
	resulting in the same welfare
	\[
		SW_{[1:T+\Gamma]}(x) \le
		\frac{7}{4}T+o(T).
	\]

	The offline optimum schedules $B$ in every first-half block and $(A,C)$ in every second-half block for a total value of $k+3k=2T$, giving the same competitive ratio $\tfrac{7}{8}+o(1)$.

	\paragraph{Conclusion.}
	In both cases the competitive ratio is at most $\tfrac{7}{8}+o(1)$, so every online fractional algorithm forfeits at least $1/8$ of the attainable social welfare.
\end{proof}

\subsection{Myopically Reasonable Algorithms Lose Nearly Half of Welfare}

For fully patient bidders we will say that transaction $i_2$ is dominated by transaction $i_1$ if (a) $v_{i_1} > v_{i_2}$ and (b) for every $j$
we have that $w_{i_1 j} \le w_{i_2j}$.

\begin{definition}
	A possibly fractional allocation $x=\{x_i^t\}_{i,t}$ is called \emph{myopically reasonable} if
	for every transaction $i_2$ that
	is dominated by some transaction $i_1$, it is
	never scheduled before $i_1$ is exhausted. I.e.,
	$x_{i_2}^t>0$ implies that $\sum_{\tau=1}^t x_{i_1}^\tau = 1$. An online algorithm
	is called myopically reasonable if it always produces myopically reasonable allocations.
\end{definition}

\begin{theorem} \label{thm:myopic-lower-bound}
	Let \( m \ge 2 \). Fix a myopically reasonable
	fractional scheduling algorithm which satisfies average block size constraint \( \{B_j\}_{j=1}^m \) with slackness \( \Delta(T) = o(T) \). Then for every extension \( \Gamma(T) = o(T) \) and horizon $T$ there exists an input with fully patient bidders such that the output of the algorithm \( x = \{x^t_i\}_{i,t} \) obtains at most a fraction of $1/2 + O(1/m)$ of the optimal social welfare, i.e.,
	\[
		\frac{SW_{[1:T+\Gamma]}(x)}{SW_{[1:T]}(y)} \le \frac{m}{2m-1}+o(1),
	\]
	where \( y=\{y_i^t\}_{i,t} \) is the optimal integral allocation (for horizon $T$) with per-block limits \( \{B_j\}_{j=1}^m \).
\end{theorem}

\begin{proof}
	Fix a horizon \(T\) and a small constant \(\varepsilon>0\).
	W.l.o.g.\ scale all resources to have unit capacity,
	i.e.,\ set \(B_j=1\) for every \(j\).

	\paragraph{Adversarial input.}
	We define an adversarial input that forces any myopically reasonable algorithm to a sub-optimal allocation.
	Let \(e_j\) denote the \(j\)-th standard basis vector in \(\mathbb{R}^m\).
	For every \(1\le k\le m\) define two transaction types:
	\[
		\text{light }L_k : \;
		\bigl(v,w\bigr)=\bigl(1+2k\varepsilon,\;e_k\bigr),
		\qquad
		\text{heavy }H_k : \;
		\bigl(v,w\bigr)=\bigl(1+(2k-1)\varepsilon,\;\sum_{j=k}^{m}e_j\bigr).
	\]
	Partition the horizon into \(m\) equal \emph{batches} of length \(L\),
	where \(T=mL\).
	Let \(t_k:=(k-1)L+1\) be the first block of batch~\(k\).
	At the beginning of batch~\(k\) the adversary releases
	\[
		R \;:=\; L+\Delta(L)+1
	\]
	transactions of each type \(L_k\) and \(H_k\).

	\paragraph{Upper-bounding the algorithm's welfare.}
	For indices \(k_1\ge k_2\) we have the strict dominance
	\[
		v(L_{k_1}) > v(H_{k_2}),
		\quad
		w(L_{k_1}) \le w(H_{k_2}),
	\]
	so a myopically reasonable algorithm never schedules a heavy
	\(H_{k_2}\) while any light \(L_{k_1}\) remains.
	Feasibility within batch \(k\) therefore yields
	\[
		\sum_{t=t_k}^{t_k+L-1}\sum_{i\in L_k} x_i^t
		\;\le\;
		\sum_{t=t_k}^{t_k+L-1}\sum_{i} w_{i,k}x_i^t
		\;\le\;
		L+\Delta(L)
		\;<\;
		R,
	\]
	hence no heavy transaction of \emph{any} type can be placed in
	blocks \(1,\dots,T\).
	Using feasibility in the extension window and the fact that every heavy
	transaction consumes resource~\(m\),
	\[
		\sum_{t=T+1}^{T+\Gamma}\sum_{i\in\bigcup_k H_k} x_i^t
		\;\le\;
		\sum_{t=T+1}^{T+\Gamma}\sum_{i} w_{i,m}x_i^t
		\;\le\;
		\Gamma+\Delta(\Gamma).
	\]
	Consequently, by bounding the contribution of each transaction type to social welfare, we get:
	\begin{align*}
		SW_{[1:T+\Gamma]}(x)
		 & \;\le\;
		\sum_{k=1}^m
		\sum_{t=1}^{T+\Gamma}\sum_{i\in L_k} x_i^tv(L_k) + \sum_{t=T+1}^{T+\Gamma}\sum_{i\in \bigcup_k H_k}x_i^tv_i
		\\&\le
		m(1+2m\varepsilon)\bigl(L+\Delta(L)+1\bigr)
		+ (1+2m\varepsilon)\bigl(\Gamma+\Delta(\Gamma)\bigr)
		\\&=
		m(1+2m\varepsilon)\bigl(L+o(T)\bigr).
	\end{align*}

	\paragraph{Optimal offline benchmark.}
	The offline integral scheduler proceeds as follows.
	During the first batch (blocks \(1,\dots,L\)) it places exactly one heavy
	transaction \(H_1\) in every block.
	In each subsequent batch \(k\ge2\) it packs the block with the pair
	\(L_{k-1}\) and \(H_k\).
	The total welfare of this allocation is
	\[
		SW_{[1:T]}(y)
		\;=\;
		\left(\,\sum_{k=1}^{m} v(H_k)
		+\sum_{k=1}^{m-1} v(L_k)\right)L
		\;\ge\;
		(2m-1)L.
	\]

	\paragraph{Competitive ratio.}
	Combining the bounds gives
	\[
		\frac{SW_{[1:T+\Gamma]}(x)}{SW_{[1:T]}(y)}
		\;\le\;
		\frac{m(1+2m\varepsilon)\bigl(L+o(T)\bigr)}
		{(2m-1)L}
		\;=\;
		\frac{m}{2m-1}\bigl(1+2m\varepsilon\bigr)+o(1).
	\]
	Choosing \(\varepsilon=o(1)\) completes the proof.
\end{proof}

\section{EIP-1559} \label{sec:1559}

As outlined in the introduction, the multidimensional EIP-1559 mechanism combines a protocol-enforced pricing rule with a block builder's packing decisions. Because current blockchain fee mechanisms do not provide a way for transactions to specify time sensitivity, we restrict our analysis in this section strictly to fully patient bidders. Furthermore, to isolate the fundamental algorithmic properties of the system, we abstract away the strategic \enquote{tip} auction and focus purely on the multidimensional block packing problem.

To formally evaluate the social welfare achieved by this mechanism, we must explicitly decouple it into two distinct algorithmic entities:

\begin{itemize}
    \item \textbf{The Protocol (\Cref{alg:1559}):} Regulates the economic environment. It is parameterized by a target capacity $B_j$ and a capacity multiplier $c_j$ for each resource $j$, which caps the maximum block capacity at $c_j B_j$. It dictates transaction validity by tracking base prices, governed by a global adjustment parameter $\eta$, an initial price $p_1$, and a minimum price floor $p_{\min}$.
    \item \textbf{The Block Builder (\Cref{alg:1559-builder}):} Determines the actual allocation. The builder is free to select any feasible subset of transactions deemed valid by the protocol. While \enquote{usually} the maximum block size is sufficient to include all valid transactions, \enquote{occasionally} it is not, a scenario our lower bounds suggest may be highly restrictive in multidimensional settings. In these bottleneck events, the mechanism relies entirely on the builder's algorithm to decide which subset to choose.
\end{itemize}

\begin{algorithm}[H]
    \caption{The Multidimensional EIP-1559 Algorithm}
    \label{alg:1559}
    
    \SetKwInOut{Input}{Input}
    \SetKwInOut{Parameters}{Parameters}
    \SetKwInOut{Uses}{Uses}
    \SetKwInOut{Output}{Output}
    \Input{Transaction set $\{(a_i, v_i, \{w_{ij}\}_{j=1}^m)\}_{i \in \mathcal{C}}$}
    \Parameters{$(\{B_j\}_{j=1}^m, \{c_j\}_{j=1}^m, \eta, p_1, p_{\min})$ where $\eta > 0$ and $p_1 > p_{\min} > 0$}
    \Uses{A block building algorithm BUILD}
    \Output{Allocation $\{x_i^t\}_{i \in \mathcal{C},\, t \in \mathbb{N}}$ with $x_i^t \in \{0,1\}$}

    \BlankLine
    Set the prices for block $t=1$ to $p_j^1=p_1$ for all $1 \le j \le m$\;
    
    \For{each time step $t = 1, 2, \dots$}{
        Let the set of valid transactions $V_t \subseteq \mathcal{C}$ be those that have arrived but not yet scheduled: $a_i \le t, \forall s < t: x_i^s = 0$, and are willing to pay current prices $v_i \ge \sum_{j=1}^m w_{ij}p_j^t$\;
        
        Call BUILD on the set of valid transactions, generating allocations $x_i^t$ for all $i$\;
        
        Update the prices for the next block for each resource $j$:
        \[
        p_j^{t + 1} = \max\left\{p_j^t \cdot e^{\eta(B_j^t/B_j - 1)}, p_{\min}\right\}
        \]
        where $B_j^t$ is the demand for resource $j$ in block $t$: $B_j^t = \sum_{i \in \mathcal{C}}w_{ij}x_i^t$\;
    }
\end{algorithm}

\begin{algorithm}[H]
    \caption{The Block Building Algorithm (BUILD)}
    \label{alg:1559-builder}
    
    \SetKwInOut{Input}{Input}
    \SetKwInOut{Parameters}{Parameters}
    \SetKwInOut{Output}{Output}
    \SetKwInOut{Implementation}{Implementation}

    \Input{Valid transaction set $\{(v_i, \{w_{ij}\}_{j=1}^m)\}_{i \in V_t}$ where $V_t\subseteq \mathcal C$, Prices $\{p_j^t\}_{j=1}^m$}
    \Parameters{Target resource usage $\{B_j\}_{j=1}^m$; Maximal capacity constraints $\{c_j\}_{j=1}^m$}
    \Output{Feasible allocation $\{x_i^t\}_{i \in V_t}$ satisfying $\sum_{i \in V_t} w_{ij}x_i^t \le c_jB_j$ for all $j$}

    \Implementation{Left to the (strategic) discretion of the block builder}
\end{algorithm}

\subsection{Feasibility of Multidimensional EIP-1559}

Before analyzing the optimality of EIP-1559, we show that it produces feasible allocations, i.e., allocations with an average block size at most $B_j$ for each resource $j$ with some constant slackness.

\begin{proposition}\label{prop:1559-feasibility}
    For every block building algorithm BUILD and input transaction sequence $\mathcal C$, the allocation produced by the EIP-1559 algorithm (\Cref{alg:1559}) has an average block size at most $B_j$ with slackness $\Delta=\frac{1}{\eta}\ln\left(\frac{V_{\max}}{p_{\min}B_j}\right) + c_j - 1$ for every resource $j$, where $V_{\max}$ is an upper bound on the social welfare obtainable by any single block, i.e., $\sum_iv_ix_i^t \le V_{\max}$ for every block $t$.
\end{proposition}

This feasibility result is derived from the following lemma, which establishes a strict upper limit on the protocol's prices.

\begin{lemma} \label{lem:price-upper-bound}
    For every block building algorithm BUILD and input sequence $\mathcal C$, if the social welfare of each block $t$ generated by the algorithm is bounded by $\sum_iv_ix_i^t \le V_{\max}$, then the prices generated by the algorithm satisfy $p_j^t \le e^{\eta(c_j-1)}V_{\max}/B_j$ for every block $t$ and resource $j$.
\end{lemma}

\begin{proof}
    Fix a resource $j$. Assume for contradiction that $p_j^t > e^{\eta(c_j-1)}V_{\max}/B_j$ for some block $t$. Let $t+1$ be the first such block.
    Since EIP-1559 guarantees that all transactions included in block $t$ yield more value than their cost (defined for prices $p^t$), the total cost is bounded from above by the total welfare, i.e.,
    \[
    \sum_{k=1}^m B_k^tp_k^t = \sum_{k=1}^m p_k^t\sum_{i\in\mathcal C}w_{ik}x_i^t = \sum_{i\in\mathcal C} x_i^t \sum_{k=1}^m p_k^t w_{ik} \le \sum_{i\in\mathcal C}v_ix_i^t.
    \]
    Thus, we get $B_j^tp_j^t \le V_{\max}$.
    However, from the prices update rule, the increase in the price of resource $j$ from block $t$ to block $t+1$ is at most $e^{\eta(c_j-1)}$, i.e., $p_j^t \ge p_j^{t+1}e^{-\eta(c_j-1)} > V_{\max}/B_j$, and we get a contradiction.
\end{proof}

\begin{proof}[Proof of \Cref{prop:1559-feasibility}]
    From the price update rule, $p_j^{t+1} \ge p_j^t e^{\eta(B_j^t/B_j - 1)}$ for every time $t$ and resource $j$. Thus, $B_j^t/B_j \le \frac{1}{\eta}\ln(p_j^{t+1}/p_j^t) + 1$.
    Summing over all blocks in the range $[T_1,T_2]$, we get:
    \[
    \sum_{t=T_1}^{T_2}B_j^t/B_j \le T_2 - T_1 + 1 + \frac{1}{\eta}\sum_{t=T_1}^{T_2}\ln(p_j^{t+1}/p_j^t) = 
    T_2 - T_1 + 1 + \frac{1}{\eta}\ln(p_j^{T_2+1}/p_j^{T_1}).
    \]
    From \Cref{lem:price-upper-bound}, we get:
    \[
    \ln(p_j^{T_2+1}/p_j^{T_1}) \le \ln\left(\frac{e^{\eta(c_j-1)}V_{\max}}{p_{\min}B_j}\right),
    \]
    thus,
    \[
    \sum_{t=T_1}^{T_2}\sum_iw_{ij}x_i^t \le \left(T_2 - T_1 + 1 + \frac{1}{\eta}\ln\left(\frac{V_{\max}}{p_{\min}B_j}\right) + c_j - 1\right)\cdot B_j.
    \]
\end{proof}

\subsection{EIP-1559 with Myopically Optimizing Block Builders}

In the analysis of the EIP-1559 algorithm we will need to relate
$\sum_t \sum_j p_j^t \cdot B_j^t$ to $\sum_t \sum_j p_j^t \cdot B_j$,
where $B_j^t$ is the total size of block $t$ in dimension $j$, so we
start with this analysis.

Fix some block building algorithm, some input sequence of transactions and some horizon $T$.
For each resource $j$ and a real number $\theta \ge p_{\min}$ denote the total number of blocks with price at least $\theta$ by 
$N_j(\theta) = |\{1 \le t \le T\mid p_j^t \ge \theta \}|$ and the total size of blocks with price at least $\theta$ by
$S_j(\theta) = \sum_{1 \le t \le T,p_j^t \ge \theta} B_j^t$.

\begin{lemma}
Let $\tau = 1 + \ln(p_1/p_{\min})/\eta$. For every resource $j$ and $\theta \ge p_{\min}$:
\begin{enumerate}
    \item if $\theta > p_1$, then $S_j\bigl(\theta e^{-\eta(c_j-1)}\bigr) \ge B_j N_j(\theta)$,
    \item if $\theta \le p_1$, then $S_j\bigl(\theta e^{-\eta(c_j-1)}\bigr) \ge B_j N_j(\theta) - B_j \tau$.
\end{enumerate}
\end{lemma}

\begin{proof}
    Fix a resource $j$. Let us represent the set of blocks with $p_j^t \ge \theta$ as a union of intervals $\bigcup_k[t^{start}_k,t^{end}_k]$. I.e., for every $t_k^{start} \le t \le t_k^{end}$ we have $p_j^t \ge \theta$ and for $t\in\{t_k^{start} - 1, t_k^{end} + 1\}$ we have $p_j^t < \theta$. 
    Fix some interval $k$ and denote $[t_k^{start},t_k^{end}] = [t_1,t_2]$. 
    
    \paragraph{Case (I) $\boldsymbol{\theta > p_1}$.} In this case, $t_1 > 1$. Since the prices of resource $j$ in the range $t\in[t_1,t_2]$ are strictly greater than $p_{\min}$, we have:
    \[
    \sum_{t=t_1 - 1}^{t_2 - 1} B_j^t = 
    \left(t_2 - t_1 + 1 + \frac{1}{\eta}\ln\frac{p_j^{t_2}}{p_j^{t_1 - 1}}\right) \cdot B_j > (t_2 - t_1 + 1) \cdot B_j
    \]

    where the inequality follows from $p_j^{t_2} \ge \theta > p_j^{t_1 - 1}$. Summing the RHS over all such intervals $[t_k^{start}, t_k^{end}]$ gives us exactly $N_j(\theta)\cdot B_j$. Since the prices decrease by at most a factor of $e^{\eta(c_j-1)}$ in a single step, we get $p_j^{t_1 - 1} \ge \theta\cdot e^{-\eta(c_j-1)}$. Thus, the interval $[t_1 - 1, t_2 - 1]$ falls within the summation that defines $S_j(\theta \cdot e^{-\eta(c_j-1)})$. Summing the inequalities over all intervals yields the claim.

    \paragraph{Case (II) $\boldsymbol{\theta \le p_1}$.} The only difference in this case is that the first block is included in the first interval. For this interval $[1,t_1^{end}]$ we derive a similar bound:
    \[
    \sum_{t=1}^{t_1^{end} - 1} B_j^t = 
    \left(t_1^{end} - 1 + \frac{1}{\eta}\ln\frac{p_j^{t_2}}{p_1}\right) \cdot B_j \ge 
    t_1^{end} \cdot B_j - \left(1 + \frac{1}{\eta}\ln\frac{p_1}{p_{\min}}\right) \cdot B_j.
    \]
    Since the same analysis used in (I) also holds for all other intervals, we get:
    \[
    S_j(\theta \cdot e^{-\eta(c_j-1)}) \ge N_j(\theta) \cdot B_j - \left(1 + \frac{1}{\eta}\ln\frac{p_1}{p_{\min}}\right) \cdot B_j
    \]
    
\end{proof}

\begin{lemma}\label{lem:1559-block-sizes}
    Let $\tau = 1 + \ln(p_1/p_{\min})/\eta$. For every resource $j$:
    \[
    e^{\eta(c_j - 1)} \sum_{t = 1}^T p_j^t B_j^t \ge 
    \sum_{t = 1}^T p_j^t B_j - B_j (p_1 \tau + p_{\min} T).
    \]
\end{lemma}

\begin{proof}
    Denote $S_j^t(\theta) = \mathds{1}_{p_j^t \ge \theta} \cdot B_j^t$. Integrating over $\theta$ we have:
    \[
    \int_0^\infty S_j^t(\theta) d\theta = B_j^t \cdot \int_0^\infty \mathds{1}_{p_j^t \ge \theta} d\theta = p_j^t B_j^t.
    \]
    Summing over all $t$ we get $\int_0^\infty S_j(\theta)d\theta = \sum_{t=1}^T p_j^t B_j^t$.
    Similarly, $B_j \cdot \int_0^\infty N_j(\theta) d\theta = \sum_{t=1}^T p_j^t B_j$. 
    
    Applying a linear change of variables, we have that:
    \[
    \int_0^\infty S_j(\theta \cdot e^{-\eta(c_j-1)})d\theta = e^{\eta(c_j-1)} \cdot \int_0^\infty S_j(\theta)d\theta.
    \]
    
    From the last lemma, for $\theta \ge p_{\min}$ it holds
    $S_j(\theta \cdot e^{-\eta(c_j-1)}) \ge \left( N_j(\theta) - \mathds{1}_{p_1 \ge \theta}\tau \right) \cdot B_j$, thus:
    \[
    e^{\eta(c_j-1)} \cdot \int_0^\infty S_j(\theta)d\theta \ge
    \int_{p_{\min}}^\infty S_j(\theta \cdot e^{-\eta(c_j-1)})d\theta \ge 
    B_j \cdot \int_{p_{\min}}^\infty (N_j(\theta) - \mathds{1}_{p_1 \ge \theta}\tau) d\theta.
    \]
    
    For $\theta \le p_{\min}$ we can use the following bound
    $\int_0^{p_{\min}} N_j(\theta) d\theta \le p_{\min} \cdot T$, and we get:
    \[
    e^{\eta(c_j-1)} \cdot \sum_{t=1}^T p_j^t B_j^t \ge 
    B_j \cdot \left(\int_0^{\infty} N_j(\theta) d\theta - \int_0^{p_{\min}} N_j(\theta) d\theta - p_1\tau\right) \ge 
    \sum_{t=1}^T p_j^t B_j - B_j(p_1 \tau + p_{\min} T),
    \]
    as claimed. 
\end{proof}

We now proceed with the analysis of the social welfare of EIP-1559. Define the class of \emph{myopically-$\lambda$-optimal} block building algorithms as follows:

\begin{algorithm}[H]
    \caption{Myopically-$\lambda$-Optimal Block Building Algorithms}
    \label{alg:myopic-builder}
    
    \SetKwInOut{Input}{Input}
    \SetKwInOut{Parameters}{Parameters}
    \SetKwInOut{Output}{Output}

    \Input{Valid transaction set $\{(v_i, \{w_{ij}\}_{j=1}^m)\}_{i \in V_t}$ where $V_t\subseteq \mathcal C$, Prices $\{p_j^t\}_{j=1}^m$}
    \Parameters{Target resource usage $\{B_j\}_{j=1}^m$; Maximal capacity constraints $\{c_j\}_{j=1}^m$}
    \Output{Feasible allocation $\{x_i^t\}_{i \in V_t}$ satisfying $\sum_{i \in V_t} w_{ij}x_i^t \le c_jB_j$ for all $j$}

    \BlankLine
    Return any (strategically-chosen) feasible allocation $\{x_i^t\}_{i\in V_t}$ which $\lambda$-approximates the optimal allocation over the valid transactions, i.e., $\sum_{i\in V_t}v_ix_i^t \ge \lambda \cdot \sum_{i\in V_t}v_iy_i^t$ for any allocation $\{y_i^t\}_{i\in V_t}$ feasible under capacity constraints $\{B_j\}_{j=1}^m$\;
\end{algorithm}

Our result shows that if the block builders manage to 
(approximately) optimize 
the current block, under the EIP-1559 restrictions of allowed transactions, then we overall get
a good approximation of the offline optimum.
Note that our algorithm is actually allowed to use size $c_j \cdot B_j$ in each dimension $j$, but we still only require it to compete with blocks that are limited to constraints $\{B_j\}_{j=1}^m$.

\EIPUpperBound*

\begin{proof}
As in the proof of our general upper bound, we separate the $y$ allocation into
the transactions that were also scheduled (at any time) by $x$, to be denoted
$y_+$, and those that were never allocated by $x$ to be denoted by $y_-$.
Clearly, we have:
\begin{equation}\label{eq-A}
    SW_{[1:T]}(x) \ge SW_{[1:T]}(y_+). 
\end{equation}

Now let us look at the allocation of $x$ at some block $t$ and compare
it to $y_-$ at block $t$. The only transactions that are allocated by $y_-$
at time $t$ and are not available to $x$ are those with $v_i < \sum_j w_{ij}p_j^t$.
The total value of these allocations is bounded from above by $\sum_j p_j^tB_j$,
thus, the optimal allocation among transactions that are allowed by EIP-1559 has value
at least $\sum_i v_iy_-^t - \sum_j p_j^tB_j$. By our assumption $x^t$ gets value
of at least $\lambda$ fraction of that, so summing up over all $t$, we get 
\begin{equation}\label{eq-B}
    SW_{[1:T]}(x) \ge \lambda \cdot \left(SW_{[1:T]}(y_-) - \sum_t \sum_j p_j^tB_j\right)
\end{equation}

Since $x^t$ only schedules
transactions with $v_i \ge \sum_j w_{ij} p_j^t$ at each block $t$, we have that the total
welfare that it obtains from block $t$ is bounded from below by $\sum_j p_j^t B_j^t$.
Summing over all $t$ we have: 
\begin{equation} \label{eq-C}
    SW_{[1:T]}(x) \ge \sum_t \sum_j p_j^t B_j^t.
\end{equation}

Let us now sum inequalities \eqref{eq-A}, \eqref{eq-B} and \eqref{eq-C}, where inequality \eqref{eq-B} is first multiplied by $1/\lambda$ and inequality \eqref{eq-C} is multiplied by $e^{\eta(c-1)}$. We get
\[
(1+1/\lambda+e^{\eta(c-1)})\cdot SW_{[1:T]}(x) \ge SW_{[1:T]}(y) - \Delta,
\]
where $\Delta = \sum_t \sum_j p_j^tB_j - e^{\eta(c-1)} \cdot \sum_t \sum_j p_j^t B_j^t$. Since $c \ge c_j$ we have $e^{\eta(c-1)} \ge e^{\eta(c_j-1)}$ and thus $\Delta \le \sum_j \left(\sum_t p_j^tB_j - e^{\eta(c_j-1)} \cdot \sum_t p_j^t B_j^t\right)$. Applying \Cref{lem:1559-block-sizes} for each $j$ gives $\Delta \le \sum_jB_j(p_1\tau + p_{\min}T)$, as needed.
\end{proof}

\subsection{Lower Bounds for EIP-1559}

Given that EIP-1559 regulates resource prices through a market-like mechanism, one might expect the system's efficiency to be robust to the specific choice of block building algorithm. Indeed, for the single-dimensional case, \cite{BN25} confirms that any builder that produces maximal blocks (i.e., does not leave \enquote{money on the table}) achieves optimal social welfare. However, we show that this guarantee collapses in the multidimensional setting. We analyze the natural \enquote{greedy} block builder, which prioritizes transactions based on their value-to-cost ratio $v_i/(\sum_j w_{ij}p_j)$. While this heuristic corresponds to the optimal LP-relaxation strategy in one dimension, we demonstrate that it fails significantly when managing multiple resources. As detailed below, this greedy approach cannot approximate the optimal welfare within a factor better than $c/m$, even under favorable conditions such as fully patient bidders and arbitrarily small transaction sizes.

\begin{algorithm}[H]
    \caption{The Greedy Block Building Algorithm}
    \label{alg:greedy-builder}
    
    \SetKwInOut{Input}{Input}
    \SetKwInOut{Parameters}{Parameters}
    \SetKwInOut{Output}{Output}

    \Input{Valid transaction set $\{(v_i, \{w_{ij}\}_{j=1}^m)\}_{i \in V_t}$ where $V_t\subseteq \mathcal C$, Prices $\{p_j^t\}_{j=1}^m$}
    \Parameters{Target resource usage $\{B_j\}_{j=1}^m$; Maximal capacity constraints $\{c_j\}_{j=1}^m$}
    \Output{Feasible allocation $\{x_i^t\}_{i \in V_t}$ satisfying $\sum_{i \in V_t} w_{ij}x_i^t \le c_jB_j$ for all $j$}

    \BlankLine
    Sort transactions $V_t$ in decreasing order of $v_i / \left(\sum_j w_{ij} p_j^t\right)$\;
    Initialize $x_i^t \leftarrow 0$ for all $i \in V_t$\;
    \ForEach{transaction $i \in V_t$ in sorted order}{
        \If{adding transaction $i$ maintains $\sum_{k \in V_t} w_{kj} x_k^t \le c_j B_j$ for all $j$}{
            $x_i^t \leftarrow 1$\;
        }
    }
\end{algorithm}

\begin{theorem}\label{thm:1559-lower-bound}
    Assume $c := \max_j c_j < m$. Then for every extension $\Gamma(T)=o(T)$ and horizon $T$ there exists an input with fully patient bidders such that the output $x=\{x_i^t\}_{i,t}$ of the EIP-1559 algorithm (\Cref{alg:1559}) that uses the greedy block building algorithm (\Cref{alg:greedy-builder}) obtains at most a factor of $c/m$ of the optimal social welfare, i.e.:
    \[
    \frac{SW_{[1:T+\Gamma]}(x)}{SW_{[1:T]}(y)} \le \frac{c}{m} + o(1),
    \]
    where $y=\{y_i^t\}_{i,t}$ is the optimal offline allocation for $T$ blocks with capacity constraints $\{B_j\}_{j=1}^m$. 
\end{theorem}

Note that \Cref{alg:greedy-builder} always produces maximal-by-inclusion allocations. Thus, as a direct corollary, we get the non-trivial result that in the multidimensional version of EIP-1559, maximal-by-inclusion allocations fail to approximate the optimal social welfare.
\footnote{One could also consider a non-skipping variant that terminates immediately upon the first resource violation, rather than iterating through the remaining transactions. The lower bound in \Cref{thm:1559-lower-bound} applies to this halting variant as well.}

\begin{proof}
Assume $c < m$. Normalize target capacities to $B_j=1$, so the hard capacity limit is $c$. We consider the \enquote{small transaction limit} where the algorithm may allocate transactions fractionally; this ensures the bound is robust against infinitesimal transaction sizes ($q_{\max} \ll 1$).

\paragraph{1. Adversarial Construction}
Fix $\epsilon > 0$. At each time $t \in [1, T+\Gamma]$, $m$ transactions $\{tx_1^t, \dots, tx_m^t\}$ arrive. Each $tx_j^t$ is defined by:
\begin{itemize}
    \item \textbf{Value:} $v_j^t = 1 + t\epsilon + r_j^t$, with unique perturbations $r_j^t \in (0, \epsilon)$.
    \item \textbf{Weights:} $w_{j,j} = c$ and $w_{j,k} = \epsilon$ for $k \neq j$.
\end{itemize}
The perturbations $r_j^t$ ensure that at any time $t$, the value-cost ratios $v_i / \sum_j w_{ij}p_j^t$ are unique, forcing a strict priority ordering.

\paragraph{2. Algorithm Performance}
Since $v_i^t$ increases with $t$, the greedy builder always prioritizes current arrivals over the backlog. However, because $w_{j,j} = c$, fully allocating any transaction $tx_j^t$ saturates resource $j$ to the limit $c$. Due to the strict ordering of value-cost ratios, the greedy builder fully exhausts the capacity of at least one resource by picking the single highest-ratio transaction $tx_j^t$ and can schedule no others in that block. Thus:
\[
SW_{[1:T+\Gamma]}(x) = \sum_{t=1}^{T+\Gamma} (1 + t\epsilon + \max_j r_j^t) \le (T+\Gamma)(1+(T+\Gamma+1)\epsilon).
\]

\paragraph{3. Optimal Welfare}
An optimal offline allocator can fractionally allocate each of the $m$ transactions at each time $t \in [1, T]$. By assigning a fraction $y_j^t = \frac{1}{c + (m-1)\epsilon}$ to every $tx_j^t$, the total load on any resource $k$ is:
\[
\sum_{j=1}^m w_{j,k} y_j^t = \frac{c + (m-1)\epsilon}{c + (m-1)\epsilon} = 1.
\]
The resulting optimal welfare is:
\[
SW_{[1:T]}(y) \ge \sum_{t=1}^T \sum_{j=1}^m \frac{v_j^t}{c+(m-1)\epsilon} \ge \frac{m T}{c + (m-1)\epsilon}.
\]

\paragraph{4. Competitive Ratio}
Combining the bounds, we have:
\[
\frac{SW_{[1:T+\Gamma]}(x)}{SW_{[1:T]}(y)} \le \frac{(T+\Gamma)(1+(T+\Gamma+1)\epsilon)}{\frac{mT}{c+(m-1)\epsilon}} = \frac{(T+\Gamma)(c+(m-1)\epsilon)(1+(T+\Gamma+1)\epsilon)}{mT}.
\]
Simplifying by grouping the $T$ terms and the $\epsilon$ terms:
\[
\frac{SW_{[1:T+\Gamma]}(x)}{SW_{[1:T]}(y)} \le \left(1 + \frac{\Gamma}{T}\right) \left(\frac{c + (m-1)\epsilon}{m}\right) (1 + (T+\Gamma+1)\epsilon).
\]
Taking the limit as $\epsilon \to 0$ and noting that $\Gamma(T) = o(T)$, the terms $(1+\Gamma/T)$ and $(1+O(T\epsilon))$ converge to 1, yielding:
\[
\frac{SW_{[1:T+\Gamma]}(x)}{SW_{[1:T]}(y)} \le \frac{c}{m} + o(1).
\]
\end{proof}

\section*{Acknowledgments}
We thank Yossi Azar, Moshe Babaioff, Uri Feige, and Seffi Naor for helpful discussions
and pointers, and Noam Manaker Morag for helpful comments on early versions of this manuscript. This work was supported by a grant from the Israel Science Foundation (ISF number 505/23).

\bibliographystyle{alpha}
\bibliography{bib}

\newpage
\appendix

\section{Approximation Algorithm for the General Setting} \label{app:general-setting}

In the general setting, achieving a computationally efficient constant-factor approximation is impossible unless P=NP. To overcome this limitation, we employ \emph{resource augmentation} as in \cite{BN25}: we allow our online algorithm to pack larger individual blocks, provided it strictly maintains an \emph{average block size} of $B_j$ (\Cref{dfn:slackness}), and permit it to run for an \emph{extension} of $\Gamma$ additional blocks.

\begin{theorem} \label{thm:general-setting-upper-bound}
	There exists an online polynomial-time algorithm that produces
	allocations of average block size at most $\{B_j\}_{j=1}^m$
	with slackness $\Delta = O(\log m)$ that approximates the optimal fractional
	allocation within a factor of $\Omega\left((1-\rho_{\max})^{O(\log m)}\right)$
	with
	$\Gamma=O(\log m)$ extension, where $\rho_{\max}=\max_i \rho_i$.
\end{theorem}

In particular, for fully patient users, i.e., $\rho_i=0$ for all $i$, we get a constant-factor approximation for the optimal welfare in the general setting. Since the slackness is $O(\log m)$, we immediately get that the maximum block size in every resource is at most an $O(\log m)$ factor larger than $B_j$.  We leave it as an open problem whether the maximum block size can be bounded by some constant multiple of $B_j$.

The idea is to use the following reduction: we batch every $L$ consecutive blocks into a single block with size larger by a factor of $L$,
leaving the intermediate blocks empty.
Batching preserves the average block capacity, incurring slackness \(L\), but scales every transaction's demand down by a factor \(1/L\) relative to the maximum block capacity.

\begin{algorithm}[H]
	\caption{The Batching Algorithm} \label{alg:batching}
	\SetKwInOut{Input}{Input}
	\SetKwInOut{Output}{Output}
	\SetKwInOut{Uses}{Uses}

	\Input{Transaction set $\{(a_i,v_i,\rho_i,\{w_{ij}\}_{j=1}^m)\}_{i\in\mathcal{C}}$; Batching parameter $L\in\mathbb{N}$}
	\Uses{Integral Block-Packing oracle $\widetilde{\text{ALG}}$}
	\Output{Integral allocation $\{x_i^t\}$}

	\For{each time step $t = 1,2,\dots$}{
		\eIf{$t \bmod L \neq 0$}{
			Return an empty block, i.e., $x_i^t = 0$ for all $i$\;
		}{
			Define batch $\tilde t=t/L$ of transactions to be
			\[
				\tilde A_{\tilde t} := \{i\in\mathcal{C}\mid t-L<a_i \le t \},
			\]
			such that each $i\in \tilde A_{\tilde t}$ has value $\tilde v_i :=v_i^t=v_i(1-\rho_i)^{t-a_i}$ and discount rate $\tilde \rho_i:=1-(1-\rho_i)^L$\;
			Compute $\tilde x_i^{\tilde t}$, the output of $\text{ALG}$ with $\tilde A_{\tilde t}$ as the input transactions at time $\tilde t$\;
			Output the allocation returned by $\text{ALG}$, i.e., $x_i^t=\tilde x_i^{\tilde t}$ for all $i$\;
		}
	}
\end{algorithm}

The batching algorithm (\Cref{alg:batching}) splits transactions into $L$-sized \emph{batches}. Each batch $\tilde t$ contains all transactions $i$ with arrival time $(\tilde t - 1) L<a_i\le \tilde tL$. The algorithm then builds $L$ times bigger blocks from the batches to effectively shrink the ratio $q_{\max}=\max_{i,j}\frac{w_{ij}}{B_jL}$ and improve the approximation using resource augmentation.

\begin{lemma} \label{lem:batching}
	Let $\widetilde{\text{ALG}}$ be an online block packing algorithm with average block size $\{\tilde{B}_j\}_{j=1}^m$ and slackness $\tilde{\Delta}$
	that $\tilde{\lambda}$-approximates the
	optimal social welfare with extension $\tilde{\Gamma}$ as long as $\tilde q_{\max} := \max_{i,j} w_{ij}/{\tilde B_j} \le \alpha$. Then the batching algorithm (\Cref{alg:batching}) that uses $\widetilde{\text{ALG}}$ is
	an online block packing algorithm with average block sizes $B_j=\tilde B_j/L$ for all $j$ and slackness
	$\Delta = (\tilde \Delta + 1) \cdot L$
	that $\lambda=(1-\rho_{\max})^{L-1}\cdot\tilde{\lambda}$-approximates the
	optimal social welfare with extension $\Gamma = (\tilde \Gamma + 1) \cdot L$
	as long as $q_{\max} := \max_{i,j} w_{ij}/B_j\le \alpha\cdot L$.
\end{lemma}

\begin{remark}
	The statement holds for approximating both the integral and the fractional optimum with an identical proof.
\end{remark}

\begin{proof}
	Fix a batching factor \(L\in\mathbb{N}\).
	Write
	\(
	\tilde t:=\lceil t/L\rceil
	\)
	for the batch index that contains block \(t\), and use a tilde to denote
	quantities handled by the oracle \(\widetilde{\mathrm{ALG}}\).

	\paragraph{1. Resource bounds (block averages and slackness).}
	Consider any sequence of \(K\) consecutive blocks
	\(T_0,\dots, T_0+K-1\).
	It spans at most
	\(
	\lceil K/L\rceil\le K/L+1
	\)
	distinct batches
	\(
	\tilde T_0,\dots,\tilde T_0+\tilde K-1
	\)
	with
	\(
	\tilde K\le K/L+1
	\).
	For every resource~\(j\),
	\[
		\sum_{t=T_0}^{T_0+K-1}\sum_i w_{ij}x_i^t
		=
		\sum_{\tilde t=\tilde T_0}^{\tilde T_0+\tilde K-1}
		\sum_i w_{ij}\tilde x_i^{\tilde t}
		\le
		(\tilde K+\tilde\Delta)\,\tilde B_j
		\le
		\bigl(K+L(\tilde\Delta+1)\bigr)\,B_j,
	\]
	since \(\tilde B_j=LB_j\).
	Thus, the algorithm has average block size
	\(B_j=\tilde B_j/L\) with slackness
	\(
	\Delta=L(\tilde\Delta+1).
	\)

	\paragraph{2. Preparing a comparison allocation for the oracle.}
	Let \(y=\{y_i^t\}\) be \emph{any} allocation that respects the original
	resource limits \(B_j\).
	Group it into batches:
	\[
		\tilde y_i^{\tilde t}
		:=
		\sum_{t=(\tilde t-1)L+1}^{\tilde tL} y_i^t .
	\]
	Because each block satisfies the capacity \(B_j\), the batch allocation satisfies the enlarged capacity \(\tilde B_j = LB_j\).
	Moreover, every transaction counted in \(\tilde y_i^{\tilde t}\) arrives no
	later than \(\tilde tL\), so \(\tilde y\) is a legal allocation of the input of \(\widetilde{\mathrm{ALG}}\).

	\paragraph{3. Guarantee provided by the oracle.}
	The ratio
	\(
	\tilde q_{\max}
	=
	\max_{i,j}w_{ij}/\tilde B_j
	=
	q_{\max}/L
	\)
	is at most \(\alpha\) by assumption, so for every horizon $\tilde T$ we have
	\begin{equation}\label{eq:oracle}
		\sum_{\tilde t=1}^{\tilde T+\tilde\Gamma}
		\sum_i \tilde x_i^{\tilde t}\,\tilde v_i^{\tilde t}
		\;\ge\;
		\tilde\lambda
		\sum_{\tilde t=1}^{\tilde T}
		\sum_i \tilde y_i^{\tilde t}\,\tilde v_i^{\tilde t}.
	\end{equation}

	\paragraph{4. Batch valuation.}
	The algorithm maps every transaction \((a_i,v_i,\rho_i,w_{ij})\) to the batched-transaction:
	\[
		\tilde a_i=\bigl\lceil a_i/L\bigr\rceil,\quad
		\tilde v_i=v_i(1-\rho_i)^{\tilde a_iL-a_i},\quad
		\tilde\rho_i=1-(1-\rho_i)^L.
	\]

	Therefore, the batch valuation of transaction $i$ in batch $\tilde t$ is given by
	\begin{align*}
		\tilde v_i^{\tilde t}
		 & =\tilde v_i(1-\tilde\rho_i)^{\tilde t-\tilde a_i}
		=v_i(1-\rho_i)^{\tilde a_iL-a_i}\,(1-\rho_i)^{L(\tilde t-\tilde a_i)}
		=v_i(1-\rho_i)^{\tilde tL-a_i}=v_i^{\tilde tL},
	\end{align*}
	which for every block $t$ in batch $\tilde t$, i.e., $(\tilde t-1)L<t\le\tilde tL$, can be bounded by
	\[
		\tilde v_i^{\tilde t}=v_i^{t}\,(1-\rho_i)^{\tilde tL-t}
		\;\ge\;
		v_i^{t}\,(1-\rho_{\max})^{L-1}.
	\]

	\paragraph{5. Summation over a matching horizon.}
	Choose \(\tilde T=\lceil T/L\rceil\).
	Because
	\(
	\lfloor(T+\Gamma)/L\rfloor
	\ge
	\tilde T+\tilde\Gamma
	\)
	for
	\(
	\Gamma=L(\tilde\Gamma+1)
	\), we get
	\[
		SW_{[1:T+\Gamma]}(x)
		=
		\sum_{t=1}^{T + \Gamma}x_i^tv_i^t
		=
		\sum_{\tilde t=1}^{\lfloor \frac{T + \Gamma}{L}\rfloor}x_i^{\tilde tL}v_i^{\tilde tL}
		\ge
		\sum_{\tilde t=1}^{\tilde T+\tilde \Gamma}\tilde x_i^{\tilde t}\tilde v_i^{\tilde t},
	\]
	and from \eqref{eq:oracle} and the bound on batch valuations,
	\[
		SW_{[1:T+\Gamma]}(x)
		\ge
		\tilde\lambda\,(1-\rho_{\max})^{L-1}
		\sum_{t=1}^{\tilde TL}\sum_i y_i^t v_i^t
		\ge
		\tilde\lambda\,(1-\rho_{\max})^{L-1}
		SW_{[1:T]}(y).
	\]

	\paragraph{6. Conclusion.}
	Because \(y\) was an arbitrary allocation obeying the original capacities,
	\Cref{alg:batching} achieves the approximation factor
	\(
	\lambda=\tilde\lambda\,(1-\rho_{\max})^{L-1}
	\)
	given extension $\Gamma=L(\tilde \Gamma + 1)$, and has average block size $B_j=\tilde B_j/L$ with slackness $\Delta=L(\tilde \Delta + 1)$, as desired.
\end{proof}

\begin{proof}[Proof of \Cref{thm:general-setting-upper-bound}] Fix a constant $\delta\in(0,1/2)$ (e.g., $\delta=1/4$).
	\paragraph{The oracle.}
	Use the algorithm ensured by \Cref{thm:rand-rounding-upper-bound} as our oracle.
	It returns, in expected
	polynomial-time, an \((\frac{1}{2}-\delta)\)-approximate \emph{integral} packing
	whenever the demand ratio satisfies
	\(
	\tilde q_{\max}\le c\,\delta^{2}/\log(m/\delta)
	\)
	and requires no slackness or extension.
	Choose
	\[
		L:=\Bigl\lceil\tfrac{\log(m/\delta)}{c\,\delta^{2}}\Bigr\rceil,
		\qquad
		\tilde B_j=B_j L \quad\text{for all }j.
	\]
	After scaling capacities by \(L\) inside each batch,
	\(
	\tilde q_{\max}=q_{\max}/L\le c\,\delta^{2}/\log(m/\delta),
	\)
	so the oracle is applicable to every batch.

	\noindent
	\textbf{Analysis via the batching lemma.}
	Invoking \Cref{lem:batching} with
	\(\tilde\lambda=\frac{1}{2}-\delta,\ \tilde\Delta=\tilde\Gamma=0\) gives
	\[
		\lambda=\left(\frac{1}{2}-\delta\right)\,(1-\rho_{\max})^{\,L-1},\qquad
		\Delta=\Gamma=L.
	\]

	Since \(L=\Theta(\log m)\) for fixed \(\delta\), we have $\lambda =\Omega\bigl((1-\rho_{\max})^{O(\log m)}\bigr)$ and $\Delta=\Gamma = O(\log m)$, as claimed.
\end{proof}

\end{document}